\documentclass[showpacs,aps,prd,nofootinbib,floatfix,amsmath,amssymb]{revtex4}
\usepackage{graphicx}

\def\beq{\begin{equation}}
\def\eeq{\end{equation}}
\def\[{\left[}
\def\]{\right]}

\def\PR{Phys. Rev.}

\def\gsim{\lower.7ex\hbox{$\;\stackrel{\textstyle>}{\sim}\;$}}
\def\lsim{\lower.7ex\hbox{$\;\stackrel{\textstyle<}{\sim}\;$}}

\begin{document}

\title{The $\gamma \gamma \rightarrow \phi_i\phi_j$ processes in
the type-III two-Higgs-doublet model}

\author{J. Hern\' andez--S\' anchez}
\email{jaimeh@ece.buap.mx} \affiliation{Fac. de Cs. de la
Electr\'onica, Benem\'erita Universidad Aut\'onoma de Puebla, Apdo. Postal 542, 72570 Puebla, Puebla, M\'exico, and Dual C-P Institute of High Energy Physics, M\'exico.}
\author{C. G. Honorato }
\affiliation{Departamento de F\'{\i}sica, CINVESTAV, Apartado Postal 14-740, 07000 M\'exico, D. F., M\'exico.}
\author{M. A. P\'erez }
\affiliation{Departamento de F\'{\i}sica, CINVESTAV, Apartado Postal 14-740, 07000 M\'exico, D. F., M\'exico.}
\author{ J. J. Toscano}
\email{jtoscano@fcfm.buap.mx}
\affiliation{Facultad de Ciencias F\'{\i}sico Matem\'aticas,
Benem\'erita Universidad Aut\'onoma de Puebla, Apartado Postal
1152, Puebla, Puebla, M\'exico.}

\date{\today}

\begin{abstract}
We discuss the implications of assuming a four-zero Yukawa texture and a general Higgs potential
 for the production of neutral Higgs boson pairs at $\gamma
\gamma$ colliders through the $\gamma \gamma \to \phi_i\phi_j$
($\phi_i = h$, $H$, $A$) reaction within the context of the  two
Higgs doublet model type III. Exact analytical expressions  for
the $\gamma \gamma \to \phi_i\phi_j$ reaction are presented. The
use of a nonlinear $R_\xi$-gauge, which considerably simplifies
the loop calculations and renders compact analytical expressions,
is stressed. We show that these processes are very sensitive to a
general structure of the Higgs potential that impact the triple
and quartic couplings of the scalar sector. We present results for
scenarios of the parameters of the model that are still consistent
with current experimental constraints. It is found that the cross
sections for the $\gamma \gamma \to \phi_i\phi_j$ processes can be
up to two orders of magnitude larger than those gotten in 2HDM
type I and type II. The possibility of a light CP--scalar is also studied.
\end{abstract}
 \maketitle

\section{Introduction}
The discovery of a Higgs boson (or several Higgs bosons) is
central  to the broad experimental programs of both the Large
Hadron Coller (LHC) and the International Linear Collider
(ILC)~\cite{ILC}, either through $e^-e^+$
collisions or via the secondary $\gamma \gamma $ and $\gamma e$
modes \cite{PPC, Tetlalmatzi:2009vt}. If there is a Higgs boson,
it is almost certain to be found at the LHC and its mass measured
by the ATLAS~\cite{ATLAS} and CMS~\cite{CMS} experiments. Then, in
the much cleaner environment of the ILC a more complete and
precise experimental analysis can be carried out to verify if it
corresponds to a Standard Model (SM) Higgs or other type of scalar
particle. The high level of complementarity between both type of
colliders has broadly been studied by diverse
groups~\cite{LHC-ILC}. Whereas LHC has a large mass reach for
direct discoveries due its high collision energy, the ILC enables
precise measurements and therefore detailed studies of direct
productions of new particles as well as high sensitivity to
indirect effects of heavier new particles. It has been found that
some processes occurring via $\gamma e$ or $\gamma\gamma$
collisions are complementary to their analogous reactions at
$e^+e^-$ collisions as the former are more appropriate to study
the bosonic sector of the SM. In particular, $\gamma \gamma$
collisions have been recognized as an invaluable tool to probe the
structure of electroweak interactions at high energies, both in
the gauge and the Higgs sectors \cite{GP}. In the gauge sector,
the reaction $\gamma \gamma \to WW$ has been widely studied,
mainly for testing any physics beyond the SM \cite{WW,ZZ}. As for
the Higgs sector, once a Higgs boson is discovered, the one-loop
process $\gamma \gamma \to H\to X$ might play an important role in
determining some properties of this elusive particle, such as
mass, total width,  CP properties, and couplings to other light
particles in a model-independent fashion \cite{H}.
Recently,  a series of papers have been published in photon-photon collisions on the more traditional types I and II of  2HDM, which spelt out the genuine phenomenological features that differentiate them from the MSSM \cite{arXiv:0903.4978} .
 In addition,
neutral particle pair production at a $\gamma\gamma$ collider can
be highly sensitive to new physics effects as processes of this
kind are naturally suppressed because they first arise at the one-loop
level, thereby providing a detailed test for the structure of
extended Higgs sectors.

 It is expected that the LHC will allow us to test the
mechanism of Electro-Weak Symmetry Breaking (EWSB),  which
represents a unique probe of a weakly-interacting theory, as is
the case of the Minimal Supersymmetric Standard Model (MSSM)
\cite{HHG} and general 2HDMs of Type I, II, X and Y (2HDM-I,
2HDM-II, 2HDM-X and 2HDM-Y)
\cite{Barger:1989fj, hep-ph/9603445, Aoki:2009ha}\footnote{ The type-X (type-Y)
2HDM is referred as the type-IV (type-III) 2HDM in Ref.
\cite{Barger:1989fj} and as the type-I' (type-II') 2HDM in Ref.
\cite{Grossman:1994jb, hep-ph/9603445}. Sometimes the most general 2HDM, in which
each fermion couples to both Higgs doublet fields, is called the
type III 2HDM\cite{Liu:1987ng}. All variants of the 2HDM were
discussed recently in \cite{HernandezSanchez:2011ti}. The version
that will be considered in this work will be called simply the
2HDM-III model.}, or whether strongly-interacting scenarios are
instead realized, like in the old Technicolor models or the ones
discussed more recently \cite{stronghix}.

The 2HDM-II has been quite attractive to date, in part because it coincides
with the Higgs sector of the MSSM, wherein each Higgs doublet couples
to the $u$- or $d$-type fermions separately. However, this is only valid at tree-level
\cite{Kanemura:2009mk,Babu-Kolda}. When radiative effects are
included, it turns out that the MSSM Higgs sector corresponds to
the most general version of the 2HDM, namely the 2HDM-III, whereby
both Higgs fields couple to both quarks and leptons. Thus, we can
consider the 2HDM-III as a generic description of physics at a
higher scale (of order TeV or maybe even higher), whose low energy
imprints are reflected in the Yukawa coupling structure. The
general 2HDM has a potential problem with flavor changing neutral
currents (FCNC) mediated by the Higgs boson, which arise when each
type of quark ($u$ and $d$) is allowed to couple to both Higgs
doublets, and FCNC could be induced at large rates that may
jeopardize the model. The possible solution to this problem of the
2HDM involves an assumption about the Yukawa structure of the
model. Then, in order to keep the FCNC  problem   under control,
we  can  choose one of the following mechanisms:  (1)
\textit{Discrete symmetries}. This choice is based on the
Glashow--Weinberg's theorem  concerning  FCNC's in models with
several Higgs doublets \cite{Glashow:1976nt}. (2)
\textit{Radiative suppression}. When a given set of Yukawa
matrices are present at tree-level, but the other ones arise only
as a radiative effect. This occurs for instance in the MSSM, where
the type-II 2HDM structure is not protected by some symmetry, and
it is transformed into a type-III 2HDM through loops-effects of
sfermions and gauginos~\cite{bakotau,myhlfvA,tsumura}. (3)
\textit{Flavor symmetries}. Suppression of FCNC effects can also
be achieved when a certain form of the Yukawa matrices that
reproduce the observed fermion masses and mixing angles is
implemented in the model, which is then named as THDM-III.  This
could be done either by implementing the Frogart-Nielsen mechanism
to generate the fermion mass hierarchies \cite{FN}, or by studying
a certain ansatz for the fermion mass matrices~\cite{fritzsch}. It
should be noted, that when this scheme is implemented, the scalar
Higgs potential must be expressed in its general form because it
is not necessary to impose a discrete symmetry.

In this paper, we will work in the context of the mechanism (3) by
implementing the so--called four-texture ansatz for Yukawa
matrices. This ansazt, is illustrated in Refs.~\cite{ourthdm3a,
DiazCruz:2009ek, GomezBock:2005hc}, in which a detailed study of
the 2HDM-III Yukawa Lagrangian was presented under the assumption
of a specific texture pattern \cite{Fritzsch:2002ga}, which
generalizes the original model of Ref.~\cite{cheng-sher}.
Phenomenological implications of the neutral Higgs sector of the
model, including Lepton Flavour Violation (LFV) and/or Flavour
Changing Neutral Currents (FCNCs) effects, have been studied
previously~\cite{ourthdm3b,tsumura}. The extension of such an
approach to investigate the charged Higgs boson phenomenology was
carried out in Refs.  \cite{DiazCruz:2009ek,
BarradasGuevara:2010xs}, in which the implications of this Yukawa
texture for the charged Higgs boson properties (masses and
couplings), as well as the resulting pattern of charged Higgs
boson decays and main production mechanisms at the LHC are
discussed. Here, we will focus on Higgs boson pair production at
$\gamma \gamma$ colliders through the  $\gamma\gamma\to
\phi_i\phi_j$ reaction,\footnote{Throughout the paper, the symbol
$\phi_i$ will stand for any of the three neutral Higgs bosons of
the 2HDM, $h,H,A$, whereas $\phi_a$ will denote exclusively a
CP-even Higgs boson.} within the context of this more general
version of the two-Higgs doublet model (2HDM-III). We will study
the three distinct modes for Higgs boson pair production that are
allowed: $\gamma \gamma \to AA$, $\gamma \gamma \to A\phi_a$, and
$\gamma \gamma \to \phi_a \phi_b$, with $\phi_a=h,H$. The impact
of the Higgs potential on the production of pairs of light Higgs
bosons, through the $\gamma \gamma \to hh$ reaction, has been
studied recently~\cite{P1,P2,P3}. The work of Ref.\cite{P1} is
focused to study the particular scenario where $h$ couples to
gauge bosons and fermions as the SM Higgs particle does, and finds
that the cross section for the $\gamma \gamma \to hh$ process can
be much larger than the SM prediction. The same scenario was
analyzed  in Ref.~\cite{P2}, where it is taken into account the
one--loop correction to the $hhh$ vertex. It was found that this
effect, together with the one--loop charged Higgs boson
contribution, produces a considerable enhancement of the cross
section. On the other hand, resonant effects due to the charged
Higgs boson and the heavy neutral one $H$ on the $\gamma \gamma
\to hh$ and $\gamma \gamma \to AA$ processes were studied
in~\cite{P3}. In this paper, we discuss the implications of
assuming a four-zero Yukawa texture and general Higgs potential in
the framework of the 2HDM-III on all the possible processes
$\gamma \gamma \to \phi_i\phi_j$, which include some scenarios of
experimental interest did not  considered in
Refs.~\cite{P1,P2,P3}. Apart from presenting analytical
expressions for the three distinct reactions, namely $\gamma
\gamma \to \phi_a \phi_b, \phi_a A, AA $, we will analyze their
unpolarized cross-sections in some scenarios that are still
consistent with the most recent bounds on the model parameters as
obtained from electroweak precision measurements~\cite{BTHDM}.
Higgs boson pair production at $\gamma\gamma$ colliders has
already been studied in the SM by means of a nonlinear
$R_\xi$-gauge~\cite{HHSM}, and in the MSSM via a linear $R_\xi$
gauge \cite{HHMSSM}. Partial results were also obtained within the
context of the 2HDM-II \cite{HHTHDM}.

On the other hand, the calculation of $\gamma\gamma \to \phi_i\phi_j$ scattering is far from trivial, mainly due to the plethora of Feynman diagrams arising in the gauge sector. Such a hard task can be significantly relieved if one uses an appropriate gauge-fixing procedure for the $W$ boson. Rather than using the conventional linear gauge-fixing procedure~\cite{LG}, we will use a nonlinear scheme that is covariant under the electromagnetic group $U_e(1)$~\cite{Fujikawa}. We will use a general renormalizable nonlinear gauge-fixing procedure for the 2HDM~\cite{NLGTHDM}, which is intended to remove the most unphysical vertices from the interaction Lagrangian, thereby facilitating the calculation of radiative corrections considerably. It turns out that this class of gauges is tailored for the calculation of $\gamma \gamma \to \phi_i \phi_j$ scattering. We will show that such a gauge not only reduces dramatically the number of Feynman diagrams, but renders manifestly gauge-invariant and ultraviolet-finite amplitudes. The relevance of nonlinear $R_\xi$ gauges~\cite{Fujikawa} to the calculation of radiative corrections has been emphasized by several authors not only within the context of the SM \cite{HHSM,NLGSM,BT}, but in some of its extensions such as the 2HDM~\cite{NLGTHDMA}, the so-called $331$ model \cite{NLG331}, and also in the model independent effective Lagrangian approach \cite{NLGEFT}.

The paper is organized as follows. In Sec. \ref{model} the main features of the Higgs-Yukawa sector of the 2HDM-III model are discussed. In Sec. \ref{R-gauge}, the main ingredients of the nonlinear $R_\xi$-gauge appropriate for this model are presented. Sec. \ref{results} is devoted to the cross-sections for $\gamma \gamma \to \phi_i\phi_j$ scattering, whereas numerical results are analyzed in Sec. \ref{discussion}. Finally, the conclusions are presented in Sec. \ref{conclusions}.


\section{The Higgs-Yukawa sector of the general Two Higss Doublet Model}
\label{model} In this section, we will discuss the main features
of the general Higgs potential and the implementation of a specific
four-zero texture in the Yukawa matrices within the 2HDM-III. When
a flavor symmetry in the Yukawa sector is implemented, discrete
symmetries in the Higgs potential are not needed, so the most
general Higgs potential must be introduced.

\subsection{The general Higgs potential in the 2HDM-III}
The 2HDM incorporates two scalar doublets of hypercharge $+1$:
$\Phi^\dag_1=(\phi^-_1,\phi_1^{0*})$ and
$\Phi^\dag_2=(\phi^-_2,\phi_2^{0*})$. The most general gauge
invariant and CP--conserving potential can be written
as~\cite{Gunion}
\begin{eqnarray}
V(\Phi_1,\Phi_2)&=&\mu^2_1(\Phi_1^\dag
\Phi_1)+\mu^2_2(\Phi^\dag_2\Phi_2)-\left(\mu^2_{12}(\Phi^\dag_1\Phi_2)+{\rm
H.c.}\right)+ \frac{1}{2}
\lambda_1(\Phi^\dag_1\Phi_1)^2+\frac{1}{2} \lambda_2(\Phi^\dag_2\Phi_2)^2+\lambda_3(\Phi_1^\dag
\Phi_1)(\Phi^\dag_2\, \Phi_2)\nonumber \\
&+&\lambda_4(\Phi^\dag_1\Phi_2)(\Phi^\dag_2\Phi_1)+
\left(\frac{1}{2} \lambda_5(\Phi^\dag_1\Phi_2)^2+\left(\lambda_6(\Phi_1^\dag
\Phi_1)+\lambda_7(\Phi^\dag_2\Phi_2)\right)(\Phi_1^\dag \Phi_2)+
{\rm H.c.}\right), \label{potential}
\end{eqnarray}
where all parameters are assumed to be real\footnote{The  $\lambda_6$ and $\lambda_7$ parameters are complex in general, but we will assume that they are real by simplicity.}. In many discussions of the 2HDM, the terms proportional to $\lambda_6$ and $\lambda_7$ are absent, as happens in the 2HDM type I and II where the  discrete symmetry $\Phi_1\to \Phi_1$ and
$\Phi_2\to -\Phi_2$ is imposed in order to avoid dangerous flavor changing
neutral current (FCNC) effects.
However, in our model where mass matrices with a four-texture are considered, it is not necessary to implement the above discrete symmetry. Therefore, we must keep the terms proportional to $\lambda_6$ and $\lambda_7$. As we will show below, these parameters play an important role in the $\gamma \gamma \to \phi_i\phi_j$ reactions. It is worth commenting that the parameters $\lambda_6$ and $\lambda_7$ are essential to obtain the decoupling limit of the model in which only one CP-even scalar is light. As long as these terms exist, there are two independent energy scales, $v$ and $\Lambda_{\rm THDM}$, and the spectrum of Higgs boson masses is such that $m_{h}$ is of the order of $v$, whereas $m_{H}$, $m_{A}$ and $m_{H^\pm}$ are all of the order of $\Lambda_{\rm THDM}$ \cite{Gunion}. In this case, all of the heavy Higgs bosons decouple in the limit of $\Lambda_{\rm THDM}\gg v$, according to the decoupling theorem. On the other hand, when the scalar potential does respect the discrete symmetry, it is impossible to have two independent energy scales \cite{Gunion}. As a consequence, all of the physical scalar masses lie on the Fermi scale $v$. Since $v$ is already fixed by the experiment, a very heavy Higgs boson can only arise through a large dimensionless
coupling constant $\lambda_i$. In this scenario the decoupling theorem is no longer valid, thereby opening the possibility for the
appearance of nondecoupling effects. In addition, since the scalar potential contains some terms that violate the $SU(2)$ custodial symmetry, nondecoupling effects can arise in one-loop induced Higgs boson couplings \cite{Kanemura}.

The scalar potential (\ref{potential}) has been diagonalized to
yield the mass-eigenstates fields. The charged components of the
doublets lead to a physical charged Higgs boson and the
pseudo-Goldstone boson associated with the $W$ gauge field:
\begin{eqnarray}
&&G^\pm_W=\phi^\pm_1c_\beta+\phi^\pm_2s_\beta,\\
&&H^\pm=-\phi^\pm_1s_\beta+\phi^\pm_2c_\beta,
\end{eqnarray}
with $\tan\beta=v_2/v_1\equiv t_\beta$, being $v_1/\sqrt{2}\;\, (v_2/\sqrt{2})$
the vacuum expectation value (VEV) associated with
$\Phi_1\,(\Phi_2)$, and
\begin{equation}
m^2_{H^\pm}=\frac{\mu^2_{12}}{s_\beta c_\beta}-\frac{1}{2} v^2 (\lambda_4+\lambda_5+t^{-1}_\beta \lambda_6+t_\beta \lambda_7),
\end{equation}
where we have introduced the shorthand notation $s_\beta=\sin\beta$ and $c_\beta=\cos \beta$. On the other hand, the imaginary part of
the neutral components $\phi^0_{iI}$ defines the neutral CP-odd
scalar and the pseudo-Goldstone boson associated with the $Z$
gauge boson. The corresponding rotation is given by
\begin{eqnarray}
&&G_Z=\phi^0_{1I}c_\beta+\phi^0_{2I}s_\beta, \\
&&A=-\phi^0_{1I}s_\beta+\phi^0_{2I}c_\beta,
\end{eqnarray}
where
\begin{equation}
m^2_{A}=m^2_{H^\pm}+\frac{1}{2} v^2(\lambda_4- \lambda_5).
\end{equation}
Finally, the real part of the neutral components of the
$\phi^0_{iR}$ doublets defines the CP-even Higgs bosons $h$ and $H$.
The mass matrix has the form:
\begin{equation}
M_{Re}=\left( \begin{array}{ccc} m_{11} & m_{12} \\
m_{12} & m_{22}\\
\end{array}\right),
\end{equation}
where
\begin{eqnarray}
&&m_{11}=m^2_{A} s^2_\beta + v^2 (\lambda_1 c_\beta^2+s^2_\beta \lambda_5+2 s_\beta c_\beta  \lambda_6),\\
&&m_{22}=m^2_{A} c^2_\beta + v^2 (\lambda_2 s_\beta^2+c^2_\beta \lambda_5+2 s_\beta c_\beta  \lambda_7), \\
&&m_{12}=-m^2_{A} s_\beta c_\beta + v^2 \Big( (\lambda_3+\lambda_4) s_\beta c_\beta+ \lambda_6 c_\beta^2+ \lambda_7 s^2_\beta \Big).
\end{eqnarray}
The physical CP-even states, $h$ and $H$, are written as
\begin{eqnarray}
&&H=\phi^0_{1R}c_\alpha+\phi^0_{2R}s_\alpha, \\
&&h=-\phi^0_{1R}s_\alpha+\phi^0_{2R}c_\alpha,
\end{eqnarray}

where

\begin{equation}
\tan 2\alpha=\frac{2m_{12}}{m_{11}-m_{22}},
\end{equation}
and
\begin{equation}
m^2_{H,h}=\frac{1}{2}\left(m_{11}+m_{22}\pm
\sqrt{(m_{11}-m_{22})^2+4m^2_{12}}\right).
\end{equation}

\subsection{The Yukawa sector in the 2HDM-III with a four-zero texture}

We shall follow Refs.~\cite{ourthdm3a, ourthdm3b}, where a
specific four-zero texture has been implemented for the Yukawa
matrices within the 2HDM-III. This allows one to express the couplings
of the neutral and charged Higgs bosons in terms of the fermion
masses, Cabibbo-Kobayashi-Maskawa
(CKM) mixing angles and certain dimensionless parameters,
which are to be bounded by current experimental constraints. The
Yukawa Lagrangian is written as follows:
\beq {\cal{L}}_{Y} =
Y^{u}_1\bar{Q}_L {\tilde \Phi_{1}} u_{R} +
                   Y^{u}_2 \bar{Q}_L {\tilde \Phi_{2}} u_{R} +
Y^{d}_1\bar{Q}_L \Phi_{1} d_{R} + Y^{d}_2 \bar{Q}_L\Phi_{2}d_{R}+ \, H. \, c.\, ,
\label{lagquarks} \eeq
\noindent where $\Phi_{1,2}=(\phi^+_{1,2},
\phi^0_{1,2})^T$ refer to the two Higgs doublets, ${\tilde
\Phi_{1,2}}=i \sigma_{2}\Phi_{1,2}^* $, $Q_{L}$ denotes the
left-handed quark doublet, $u_{R} $ and $d_{R}$ are the
right-handed quarks singlets, and $Y_{1,2}^{u,d}$ denotes the
$(3 \times 3)$ Yukawa matrices. The Yukawa lepton sector is given by a similar expression.

Since the fermionic contribution to the one--loop reactions $\gamma \gamma \to \phi_i\phi_j$ is given by vertices $f f \phi_i$ involving only neutral Higgs bosons, we will concentrate only in this parte of the Yukawa sector. After implementing the diagonalizations carried out in the Higgs potential and in the Yukawa sector\footnote{The
details of both diagonalizations are presented in
Ref.~\cite{ourthdm3a}.}, the interactions of the neutral
Higgs bosons $(h^{0}, H^{0}, A^{0})$ with quark pairs acquire the
following form:\\

\begin{eqnarray}
{\cal{L}}_Y^{q} & = &
\frac{g}{2}\left(\frac{m_{d_{i}}}{m_W}\right)
\bar{d_{i}}\left[\frac{ \, \cos\alpha}{\cos\beta}\delta_{ij}+
\frac{\sqrt{2} \, \sin(\alpha - \beta)}{g \, \cos\beta}
\left(\frac{m_W}{m_{d_{i}}}\right)(\tilde{Y}_2^d)_{ij}\right]d_{j}H^{0}
\nonumber \\
&  &+ \frac{g}{2}\left(\frac{m_{d_{i}}}{m_W}\right)\bar{d}_{i}
\left[-\frac{\sin\alpha}{\cos\beta} \delta_{ij}+ \frac{\sqrt{2} \,
\cos(\alpha - \beta)}{g \, \cos\beta}
\left(\frac{m_W}{m_{d_{i}}}\right)(\tilde{Y}_2^d)_{ij}\right]d_{j}
h^{0}
\nonumber \\
& &+ \frac{ig}{2}\left(\frac{m_{d_{i}}}{m_W}\right)\bar{d}_{i}
\left[-\tan\beta \delta_{ij}+  \frac{\sqrt{2} }{g \, \cos\beta}
\left(\frac{m_W}{m_{d_{i}}}\right)(\tilde{Y}_2^d)_{ij}\right]
\gamma^{5}d_{j} A^{0} \nonumber \\
& &+ \frac{g}{2}\left(\frac{m_{u{i}}}{m_W}\right)
\bar{u}_{i}\left[\frac{ \, \sin\alpha}{\sin\beta}\delta_{ij}-
\frac{\sqrt{2} \, \sin(\alpha - \beta)}{g \, \sin\beta}
\left(\frac{m_W}{m_{u_{i}}}\right)(\tilde{Y}_1^u)_{ij}\right]u_{j}H^{0}
\nonumber \\
&  &+ \frac{g}{2}\left(\frac{m_{u_{i}}}{m_W}\right)\bar{u}_{i}
\left[\frac{\cos\alpha}{\sin\beta} \delta_{ij}- \frac{\sqrt{2} \,
\cos(\alpha - \beta)}{g \, \sin\beta}
\left(\frac{m_W}{m_{u_{i}}}\right)(\tilde{Y}_1^u)_{ij}\right]u_{j}
h^{0}
\nonumber \\
& &+ \frac{ig}{2}\left(\frac{m_{u_{i}}}{m_W}\right)\bar{u}_{i}
\left[-\cot\beta \delta_{ij} + \frac{\sqrt{2} }{g \, \sin\beta}
\left(\frac{m_W}{m_{u_{i}}}\right)(\tilde{Y}_1^u)_{ij}\right]
\gamma^{5}u_{j} A^{0}, \label{QQH}
\end{eqnarray}

\noindent where $i=1,2,3$, with $d_{1}=d, d_{2}=s, d_{3}=b,
u_{1}=u, u_{2}=c, u_{3}=t$.  These couplings
depend on the rotated matrices $\tilde{Y}_n^{q} = V_q\, Y^{q}_n \, V_q^\dagger$
($n=1$ when $q=u$, and $n=2$ when
$q=d$ ). Here $V_q$ is the diagonalizing mass matrix.
In order to evaluate
$\tilde{Y}^{q}_n$ we need to focus in the quark mass matrix, which is given by,
\begin{equation}
 M^q = \frac{1}{\sqrt{2}}(v_1 \, Y_1^q + v_2 \, Y_2^q), \qquad
(q = u,\, d).
\end{equation}
We shall consider that all Yukawa matrices have
the Hermitian four-zero texture form~\cite{Fritzsch:2002ga}, and
the quark masses have the same form, which are given by:
\begin{equation} M^q= \left( \begin{array}{ccc}
0 & C_{q} & 0 \\
C_{q}^* & \tilde{B}_{q} & B_{q} \\
0 & B_{q}^*  & A_{q}
\end{array}\right)  \qquad
(q = u, d) .
\end{equation}
This is called a four-zero texture because one assumes that the
Yukawa matrices are Hermitian, therefore each $u$ and $d$ type
Yukawa matrix contains two independent zeros. According to current
analysis this type of texture satisfies the experimental
constraints  (i.e.  the Flavor Violating Higgs interaction) and at
the same time it permits to derive analytical expressions for the
Higgs boson fermion couplings
\cite{ourthdm3a,GomezBock:2005hc,DiazCruz:2009ek,ourthdm3b,BarradasGuevara:2010xs}.

To diagonalize these mass matrices, we use
the matrix $ V_q$ \footnote{$ V_q $ is built as a product of two matrices $O_q$ and $P_q$ in the form
$V_q =O_q^T \, P_q$. These matrices are given in~\cite{Fritzsch:2002ga}.} in the following way:\begin{equation}
\label{masa-diagonal}
\bar{M}^q = V_q \, M^{q} \,V^{\dagger}_q.
\end{equation}
Following the analysis in~\cite{ourthdm3a} one can derive a better
approximation for the product $V_q\, Y^{q}_n \, V_q^\dagger$,
expressing the rotated matrix $\tilde {Y}^q_n$, in the form
\begin{equation}
\left[ \tilde{Y}_n^{q} \right]_{ij}
= \frac{\sqrt{m^q_i m^q_j}}{v} \, \left[\tilde{\chi}_{n}^q \right]_{ij}
=\frac{\sqrt{m^q_i m^q_j}}{v}\,\left[\chi_{n}^q \right]_{ij}  \, e^{i \vartheta^q_{ij}},
\end{equation}
where $\chi$'s are unknown dimensionless parameters of the model, they come
from the election of a specific texture of the Yukawa matrices. In order to perform our phenomenological study, we find it convenient
to rewrite the Lagrangian given in Eq.~(\ref{QQH})  in terms of the
coefficients $ \left[\tilde{\chi}_{n}^q \right]_{ij}$, as follows:
\begin{eqnarray}
{\cal{L}}_Y^{q} & = & \frac{g}{2} \, \bar{d_i}
\left[\left( \, \frac{m_{d_i}}{m_W}\right)\frac{\cos\alpha}{\cos\beta} \,
\delta_{ij} + \frac{\sin(\alpha - \beta)}{\sqrt{2} \, \cos\beta}
\left(\frac{\sqrt{m_{d_i} m_{d_j}}}{m_W}\right)\tilde{\chi}_{ij}^d\right]d_jH^{0}
\nonumber \\
                &   & + \frac{g}{2} \, \bar{d_i}
\left[-\left(\frac{m_{d_i}}{m_W}\right)\frac{\sin\alpha}{\cos\beta} \,
\delta_{ij} + \frac{\cos(\alpha - \beta)}{\sqrt{2} \, \cos\beta}
\left(\frac{\sqrt{m_{d_i} m_{d_j}}}{m_W}\right)\tilde{\chi}_{ij}^d\right]d_jh^{0}
\nonumber \\
                &   & + \frac{ig}{2} \, \bar{d_i}
\left[-\left(\frac{m_{d_i}}{m_W}\right)\tan\beta \, \delta_{ij} +
\frac{1}{\sqrt{2} \, \cos\beta}
\left(\frac{\sqrt{m_{d_i} m_{d_j}}}{m_W}\right)\tilde{\chi}_{ij}^d\right]
\gamma^{5}} d_j A^{0.\nonumber \\
                &   & \frac{g}{2} \, \bar{u_i}
\left[\left( \, \frac{m_{u_i}}{m_W}\right)\frac{\sin\alpha}{\sin\beta} \,
\delta_{ij} - \frac{\sin(\alpha - \beta)}{\sqrt{2} \, \sin\beta}
\left(\frac{\sqrt{m_{u_i} m_{u_j}}}{m_W}\right)\tilde{\chi}_{ij}^u\right]u_jH^{0}
\nonumber \\
                &   & + \frac{g}{2} \, \bar{u_i}
\left[\left(\frac{m_{u_i}}{m_W}\right)\frac{\cos\alpha}{\sin\beta} \,
\delta_{ij} - \frac{\cos(\alpha - \beta)}{\sqrt{2} \, \sin\beta}
\left(\frac{\sqrt{m_{u_i} m_{u_j}}}{m_W}\right)\tilde{\chi}_{ij}^u\right]u_jh^{0}
\nonumber \\
                &   & + \frac{ig}{2} \, \bar{u_i}
\left[-\left(\frac{m_{u_i}}{m_W}\right)\cot\beta \, \delta_{ij} +
\frac{1}{\sqrt{2} \, \sin\beta}
\left(\frac{\sqrt{m_{u_i} m_{u_j}}}{m_W}\right)\tilde{\chi}_{ij}^u\right]
\gamma^{5}} u_j A^{0.
\end{eqnarray}
where we have redefined $\left[ \tilde{\chi}_{1}^u \right]_{ij} =
\tilde{\chi}^u_{ij}$ and $\left[ \tilde{\chi}_{2}^d \right]_{ij} =
\tilde{\chi}^d_{ij}$.
As it was discussed in Ref.~\cite{ourthdm3a}, most low-energy
processes imply weak bounds on the coefficients
$\tilde{\chi}^q_{ij}$, which turn out to be of $O(1)$.
Based on the analysis of $B \to X_s
\gamma$ \cite{Borzumati:1998nx, Xiao:2003ya}, we find
the bounds: $|\chi_{33}^{u,d}| \lsim 1$ for $0.1 < \tan \beta
\leq 70$ \cite{DiazCruz:2009ek}.  Other
constraints on the charged Higgs mass and $\tan\beta$,  can be obtained from anomalous magnetic moment of the muon
$\Delta a_{\mu}$, the $\rho$ parameter, as well as B-decays into the
tau lepton \cite{BowserChao:1998yp,WahabElKaffas:2007xd}. For instance, as it can
be read from Ref. \cite{Isidori:2007ed}, one has that the decay $B
\to \tau \nu$, implies a constraint such that for $m_{H^+}=200$
(300) GeV, values of $\tan\beta$ less than about 30 (50) are still
allowed, within MSSM or THDM-II. However, these constraints can
only be taken as estimates, as it is likely that they would be
modified for THDM-III. A more detailed analysis that includes the most
recent data is underway \cite{lorenzoetal, Mahmoudi:2009zx}.
On the other hand, the condition $\frac{\Gamma_{H^+}}{m_{H^+}} < \frac{1}{2}$ in the
frame of the 2HDM-II implies $\frac{\Gamma_{H^+}}{m_{H^+}} \approx
\frac{3G_F m_t^2}{4\sqrt{2}\pi\tan\beta^2}$ which leads to $0.3
\lsim \tan\beta \lsim 130$. However,  we found that in 2HDM-III
$\frac{\Gamma_{H^+}}{m_{H^+}} \approx \frac{3G_F
m_t^2}{4\sqrt{2}\pi\tan\beta^2} \bigg(
\frac{1}{1-\frac{\tilde{\chi}^u_{33}}{\sqrt{2} \cos\beta}}\bigg)^2$ \cite{DiazCruz:2009ek},
we have checked numerically that this leads to $0.08 < \tan\beta <
200$ when $|\tilde{\chi}^u_{33}| \approx 1$ and  $0.3 < \tan\beta <
130$ as long as $|\tilde{\chi}^u_{33}| \to 0$ recovering the result
for the case of the 2HDM-II \cite{Barger:1989fj, Chankowski:1999ta}.

Other important bounds on $\vert \tilde{\chi}_{33}\vert$  and
tan$\beta$ come from  radiative corrections to the  process
$\Gamma(Z \to b \bar{b})$, specially the hadronic branching
fraction of $Z$ bosons to $b\bar{b}$ ($R_b$) and the $b$ quark
asymmetry ($A_b$) impossed a high restriction
\cite{BarradasGuevara:2010xs,Haber:1999zh}. We can get bounds for
$\tan \beta$:  in the case $\chi_{33}^{u,d} = 1 $ and $m_{H^+}
\sim 200 (300)$ GeV,  the  range $\tan \beta > 0.3 (0.2)$ is
allowed, while in the scenario $\chi_{33}^{u,d} = -1 $ and
$m_{H^+} \sim 200 (300)$ GeV, $\tan \beta > 5 (3)$ is permitted.

On the other hand, the leading contribution to $ B_0-\bar{B}_0$
mixing in the regime small $\tan \beta$  is given by the charged
Higgs sector. Following the Ref.
\cite{BarradasGuevara:2010xs,BowserChao:1998yp},  we get for the
case $\chi_{33}^{u,d} = 1 $ and $m_{H^+} \sim 200 (300)$ GeV,
$\tan \beta > 0.2 (0.25)$ is allowed. Combining the criteria of
the analysis for the radiative corrections of $Z b \bar{b}$ vertex
and $ B_0-\bar{B}_0$ mixing,  $\tan \beta > 0.3$ is allowed for
$m_{H^+}
> 170 $ GeV and $ \chi_{33}^{u,d}=1$. However, when  $
\chi_{33}^{u,d}=-1$  and  $m_{H^+} <600 $ GeV, $\tan \beta < 2$ is
disfavored.

Besides, following the analysis of the Ref. \cite{HHG,
BowserChao:1998yp, BarradasGuevara:2010xs}, one can get the
deviation $\Delta \rho_0$ of the parameter $\rho_0= M_W^2/ \rho
M_Z^2 C^2_W$ of our version 2HDM-III, where the $\rho$ in the
denominator absorbs all the SM corrections, and the most important
SM correction at 1-loop level comes from the heavy top-quark.
According the reported value of $\rho_0$ is
\cite{Nakamura:2010zzi}
\begin{equation}
\label{rho0}
\rho_0 = 1.004 \stackrel{\scriptstyle +0.0027}{\scriptstyle -0.0007}
\;\; (2\sigma)\,.
\end{equation}
In terms of new physics (2HDM-III) the constraint becomes:
\begin{equation}
-0.0007 < \Delta \rho_{\rm 2HDM-III} < 0.0027 \;.
\end{equation}
In 2HDM $\rho_0$ receives contribution from the Higgs bosons given
by, in the context of model III \cite{HHG,
BowserChao:1998yp}
\begin{eqnarray}
\label{rho}
\Delta \rho_{\rm 2HDM-III} &=& \frac{G_F}{8\sqrt{2}\pi^2} \biggr [
\sin^2 (\alpha-\beta) F(M_{H^\pm}, M_A, M_{H^0}) \nonumber \\
&+&\cos^2(\alpha-\beta) F(M_{H^\pm}, M_A,
 M_{h^0})
\biggr] \,,
\end{eqnarray}
where
\begin{eqnarray}
F(m_1,m_2,m_3) &=& m_1^2 -\frac{m_1^2 m_2^2}{m_1^2 - m_2^2}
 \log\left( \frac{m_1^2}{m_2^2} \right) \nonumber \\
&& -\frac{m_1^2 m_3^2}{m_1^2 - m_3^2}
 \log\left( \frac{m_1^2}{m_3^2} \right)
+\frac{m_2^2 m_3^2}{m_2^2 - m_3^2}
 \log\left( \frac{m_3^2}{m_3^2} \right) \nonumber \;.
\end{eqnarray}
Since $\rho_0$ is constrained to be around 1 we have to minimize
the contributions of $\Delta\rho_{\rm 2HDM-III}$. This is obtained
for the case $\alpha = 0, \pi/2$, and the parameter space of the
scalar sector is strongly reduced when  decoupling between Higgs
bosons, i.e. $\Delta m_{ij}= m_i-m_j > 100$ GeV ($m_i= m_{h^0},
\,m_{H^0}, \, m_{A^0}, \,m_{H^\pm}$). However, is possible to
avoid the constraint for $\Delta \rho_{\rm 2HDM-III} $ if  the
decoupling  source $\Delta m_{ij} \sim 20$ GeV or $\Delta m_{ij}
\sim 100$ GeV and one Higgs very heavy (e.g. $m_{H^0}> 1$ TeV).
When $\alpha = \beta \pm \pi/2$ the allowed parameter region is
larger and one can avoid the constraints of the $\rho$ parameter
with or without decoupling. Another interesting possibility is when we have the case
quasi-degenerate  between the masses of CP-even Higgs boson ($H$) and the charged Higgs
boson ($H^\pm$). The reason is that, in a 2HDM, the custodial symmetry may be
implemented either with $m_A =m_{H^\pm}$
or with  $m_H = m_{H^\pm}$ \cite{hep-ph/0703051}.
The full study of the $\rho$ parameter
in our version 2HDM-III will be presented elsewhere
\cite{lorenzoetal}.

Hereafter, we shall refer to three benchmark  scenarios, namely:

 \begin{itemize}

\item  {\bf Scenario I (the decoupling limit)}. In this scenario, $h$ assumes the role of the SM Higgs boson  $h_{SM}$ and is essentially independent from the diverse versions of the model. We have chosen to discuss the $\gamma \gamma \to hh$ process within this context to illustrate the decoupling nature of the heavy Higgs effects~\cite{Gunion}. We will take $m_h = 120$ GeV and $m_A \sim m_{H^\pm} \sim m_H >> v$. Two cases will be considered, one when the parameter of the Higgs potential $\mu_{12}$ is of the order of the Fermi scale, $\mu_{12}\sim v$, and other when this parameter is much larger that such scale, $\mu_{12} >> v$.

 \item {\bf Scenario II ( SM-like)}  \cite{Gunion}. This is a scenario of the 2HDM-III in which the couplings $hVV$ ($V=W,Z$), $hhh$, $hhhh$ are nearly indistinguishable from the corresponding $h_{SM}$, whereas the $hf\bar{f}$ couplings can deviate significantly from the corresponding $h_{SM} f \bar{f}$ ones. We will take the following values for the  parameters of the model $m_h = 120$ GeV, $m_A= 110$ GeV, $\mu_{12}=130$ GeV, $  m_{H^\pm} \sim m_H \sim m_A+m_h$ GeV, and $\alpha=\beta\pm \pi/2$. Within the two subscenarios: $(\lambda_7= -\lambda_6=-0.1)$  and $(\lambda_7= -\lambda_6=-1)$.

 \item {\bf Scenario III (a more general case of 2HDM-III)}.  In this scenario, more general couplings of neutral Higgs bosons to SM particles are assumed. We will include the contributions of the parameters of the Higgs potential $\lambda_6$ and $\lambda_7$, as well as the contributions of the Yukawa texture in the couplings $\phi f \bar{f}$. Within this scenario, a degenerate case, a nondegenerate case and the case with a light CP-odd scalar  will be considered. In the nondegenerate case, we will choose  $m_{H^\pm} =400$ GeV, $m_A=350$ GeV, $m_H=520$ GeV, $m_{h} =120$ GeV, and $\mu_{12}=120$ GeV. On the other side, in the degenerate case we will choose $m_{H^\pm}=m_H=m_A=300$ GeV, with $\mu_{12}=60$ GeV and $m_h=120$ GeV.  For the case with light CP-odd scalar, we will choose   $m_A = 50 $ GeV,  $m_{h} =120$ GeV, $m_{H^\pm} =350$ GeV, $m_H=400$ GeV, and $\mu_{12}=70$ GeV.
 In all cases, the $\alpha=\beta$ and $\alpha=\beta\pm \pi/2$ possibilities will be considered. In the nondegenerate case, the set of values $\lambda_7= -\lambda_6=-1$ and $\lambda_7= -\lambda_6=-0.1$ will be considered. As we will see below, only the values $\lambda_7= -\lambda_6=-1$ are relevant for the degenerate case and when we study the case with a light CP-odd scalar.

 \end{itemize}

 In all the above scenarios, we consider the constraints imposed by
perturbativity, $Z \rightarrow b \bar{b}$, $\rho_0$ parameter, and $B^0-\bar{B}^0$ mixing. Our predictions will be consistent with current bounds on the charged Higgs mass obtained at Tevatron~\cite{Abulencia:2005jd} and LEP2~ \cite{lepbounds,Nakamura:2010zzi}, as well as with those derived theoretically~\cite{unitarity}.


\section{The gauge-fixing procedure}
\label{R-gauge}
As already mentioned, in calculating the
$\gamma \gamma \to \phi_i \phi_j$  reaction we will define the $W$
gauge boson propagator using a nonlinear gauge--fixing procedure
that is covariant under the electromagnetic gauge group and
consistent with renormalization theory. The details of this
gauge--fixing procedure for the 2HDM has been reported recently
in~\cite{NLGTHDM}. Here, we present the gauge--fixing functions,
including some results and comments that are needed in calculating
the amplitude for the $\gamma \gamma \to \phi_i \phi_j$ process.

To begin with, we discuss the most general structure of
gauge-fixing  functions $f^a$ and $f$ for the $SU_L(2)$ and
$U_Y(1)$ gauge groups that are allowed by  renormalization theory
and covariance under the electromagnetic gauge group in the
context of the 2HDM. As it is stressed in Ref.\cite{NLGTHDM}, our main aim is to remove the most nonphysical
vertices that are generated by the Higgs kinetic-energy term. This is achieved by introducing the following
nonlinear gauge-fixing functions~\cite{NLGTHDM}:
\begin{eqnarray}
&&f^a=f^a_V+f^a_S, \\
&&f=f_V+f_S,
\end{eqnarray}
where
\begin{equation}
f^a_V=\Big(\delta^{ab}\partial_\mu-g'\epsilon^{3ab}B_\mu\Big)W^{b\mu},
\end{equation}
\begin{eqnarray}
f^a_S=\frac{ig\xi}{2}\Big\{&&\sum_{i=1}^2\Big[\Phi^\dag_i(\sigma^a-i\epsilon^{3ab}\sigma^b)\Phi_{0i}-
\Phi^\dag_{0i}(\sigma^a+i\epsilon^{3ab}\sigma^b)\Phi_i\Big]\nonumber \\
&&+i\epsilon^{3ab}(c_\beta\Phi^\dag_1+s_\beta\Phi^\dag_2)\sigma^b(c_\beta
\Phi_1+s_\beta \Phi_2)\Big\},
\end{eqnarray}
and
\begin{eqnarray}
&&f_V=\partial_\mu B^\mu, \\
&&f_S=\frac{ig'\xi}{2}\sum_{i=1}^2\Big(\Phi^\dag_i\Phi_{0i}-\Phi^\dag_{0i}\Phi_i\Big).
\end{eqnarray}
In the above expressions, $\Phi^\dag_{0i}=(0,v_i/\sqrt{2})$,
$\sigma^a$ are the Pauli matrices, and $W^a_\mu$ and $B_\mu$ are the
gauge fields associated with the electroweak group. Our gauge-fixing
functions contain the conventional linear functions as a particular
case, which are obtained when $\epsilon^{3ab}$ is set to
zero. Also, it is worth mentioning that this
gauge-fixing procedure contains as a particular case an analogous
gauge scheme for the minimal SM \cite{BT}, which becomes evident
when the $\Phi_1$ doublet is associated with the SM one and $\beta$
is set to zero.

To fully appreciate the structure of the gauge-fixing functions,
it is convenient to express them in terms of mass eigenstates
fields. After this, one obtains for the vector sector
\begin{eqnarray}
&&f^+_V=\bar{D}_\mu W^{+\mu}, \\
&&f^Z_V=\partial_\mu Z^\mu, \\
&&f^A_V=\partial_\mu A^\mu,
\end{eqnarray}
and for the scalar sector
\begin{eqnarray}
&&f^+_S=-\frac{ig\xi}{2}\Big(\varphi^0-iG_Z\Big)G^+_W,
\\
&&f^Z_S=-\xi m_ZG_Z, \\
&&f^A_S=0,
\end{eqnarray}
where $\varphi^0=v+c_{\beta-\alpha}H+s_{\beta-\alpha}h$ and
$\bar{D}_\mu=\partial_\mu-ig'B_\mu$, being $g'$ the coupling
constant associated with the $U_Y(1)$ group. We can see that both
$f^+_V$ and $f^+_S$ are nonlinear and transform covariantly under
the $U_e(1)$ group, as $\bar{D}_\mu$ contains the covariant
derivative associated with this group.

Following the study of Ref.~\cite{NLGTHDM}, the gauge-fixing Lagrangian, ${\cal L}_B$, can then be written as
\begin{equation}
{\cal L}_B={\cal L}_{BV}+{\cal L}_{BS}+{\cal L}_{BSV},
\end{equation}
where
\begin{equation}
{\cal
L}_{BV}=-\frac{1}{\xi}(\bar{D}_\mu W^{+\mu})^\dag (\bar{D}_\nu
W^{+\nu})-\frac{1}{2\xi}(\partial_\mu
Z^\mu)^2-\frac{1}{2\xi}(\partial_\mu A^\mu)^2,
\end{equation}
\begin{equation}
{\cal L}_{BS}=-\frac{g^2\xi}{4}\Big(\varphi^{2}+G^2_Z\Big)G^-_WG^+_W
-\frac{1}{2}\xi m^2_ZG^2_Z,
\end{equation}
\begin{equation}
{\cal
L}_{BSV}=\frac{ig}{2}\Big[(\bar{D}_\mu W^{+\mu})^\dag
(\varphi-iG_Z)G^+_W-(\bar{D}_\mu
W^{+\mu})(\varphi+iG_Z)G^-_W\Big]+m_ZG_Z\partial_\mu Z^\mu.
\end{equation}
We
restrict our discussion to present some comments concerning the
impact of the ${\cal L}_{BV}$, ${\cal L}_{BS}$, and ${\cal
L}_{BSV}$ Lagrangians on the Yang--Mills, the Higgs
kinetic--energy, and the Higgs potential sectors, respectively.
First of all, the term ${\cal L}_{BV}$ defines the propagators of
the gauge fields and also modifies nontrivially the Lorentz
structure of the trilinear and quartic vertices arising from the
Yang-Mills sector. Indeed, with the exception of the $WWWW$
vertex, all trilinear and quartic vertices are modified by the
gauge fixing procedure. Since the term that introduces the
modifications on these vertices is invariant under the $U_e(1)$
group~\cite{NLGTHDM}, the trilinear electromagnetic vertices
satisfy QED-like Ward identities. This fact is relevant for
radiative corrections as such a symmetry greatly simplifies this
class of calculations.

As for the ${\cal L}_{BS}$ term, it defines the masses of the
$G_W$ and $G_Z$ fields and modifies some nonphysical couplings
arising from the Higgs potential. In this gauge, the couplings
between scalar fields arise solely from the sum of the following
terms $-V(\Phi_1,\Phi_2)+{\cal L}_{BS}$. The last term in this sum
leads  to modifications in the strength of the nonphysical
couplings $HG^-_WG^+_W$, $hG^-_WG^+_W$, $H^{2}G^-_WG^+_W$,
$h^{2}G^-_WG^+_W$, $HhG^-_WG^+_W$, and $G^2_ZG^-_WG^+_W$. The
physical couplings remain unchanged, as required.

On the other hand, the term ${\cal L}_{BSV}$ considerably affects
the Higgs kinetic-energy sector of the theory since it removes
several nonphysical vertices. When these two terms are combined,
one finds~\cite{NLGTHDM} that not  only the the mixing terms
$W-G_W$ and $Z-G_Z$ are removed from the theory, as it occurs in
conventional linear gauges, but also the nonphysical vertices
$WG_W\gamma$, $WG_WZ$, $HWG_W\gamma$, $hWG_W\gamma$,
$G_ZWG_W\gamma$, $HWG_WZ$, $hWG_WZ$, and $G_ZWG_WZ$ are removed.
In addition,  the unphysical vertices $HWG_W$, $hWG_W$, and
$G_ZWG_W$ are modified. Once again, it should be emphasized that
the couplings involving only physical scalars are not modified by
the gauge-fixing procedure.

As far as the ghost sector is concerned, it is shown in
Ref.~\cite{NLGTHDM} that it shows new interesting aspects that are
not present in conventional linear gauges, such as manifest
electromagnetic gauge invariance and the presence of quartic ghost
interactions. As a consequence of $U_e(1)$--gauge invariance, the
corresponding electromagnetic couplings satisfy QED--like Ward
identities, which considerably simplifies the loop calculations
associated with the $\gamma \gamma \to \phi_i\phi_j$ process.

The gauge--dependent Feynman rules necessary for the calculation of the $\gamma
\gamma \to \phi_i \phi_j$ processes are not presented here, as they are given in Ref.~\cite{NLGTHDM}. The rest of Feynman rules, which do not depend of the gauge--fixing procedure, are given in an Appendix \ref{a1}.
\section{The processes $\gamma \gamma
\to \phi_i \phi_j$} \label{results} We now will exploit the
nonlinear $R_\xi$--gauge already introduced to calculate the $\gamma
\gamma \to \phi_i \phi_j$ process. To begin with, we would like to
discuss the basics of this process, such as its kinematics and the
gauge structure dictated by electromagnetic gauge invariance. To
this end, we use the following notation:
\begin{equation}
A_\mu (k_1)+A_\nu(k_2)\to \phi_i(k_3)+\phi_j(k_4),
\end{equation}
where the particle momenta satisfy the kinematic relation
$k_1+k_2=k_3+k_4$. The Mandelstam variables associated with this
process are $s=(k_1+k_2)^2$, $t=(k_2-k_3)^2$, and $u=(k_1-k_3)^2$,
which fulfill the relationship $s+t+u=k^2_3+k^2_4$. One useful
quantity is the transversal momentum, given by
\begin{equation}
k^2_T=\frac{1}{s}(tu-k^2_3k^2_4).
\end{equation}
As far as the gauge structures are concerned, two possibilities
arise depending on the CP properties of the final particles. One
possibility corresponds to final states with two Higgs particles
both CP--even, i.e. $\phi_a \phi_a$ or both CP--odd, $AA$.
Although it is possible to construct at least three gauge
electromagnetic structures, only two of them are independent. We
find it convenient to use the following basis:
\begin{eqnarray}
&&P_{1\mu \nu}=\frac{\sqrt{2}}{s}\Big(k_{2\mu}k_{1\nu}-k_1\cdot
k_2 g_{\mu \nu}\Big), \\
&&P_{2\mu \nu}=\frac{\sqrt{2}}{k^2_Ts}\Big[\frac{1}{2}k^2_Tsg_{\mu
\nu}+k^2_3k_{2\mu}k_{1\nu}-2k_2\cdot k_3
k_{3\mu}k_{1\nu}-2k_1\cdot k_3k_{2\mu}k_{3\nu}+2k_1\cdot
k_2k_{3\mu}k_{3\nu}\Big],
\end{eqnarray}
which is orthonormal in the sense that $P^{\mu \nu}_iP_{j\mu
\nu}=\delta_{ij}$. Another possibility corresponds to final
particles with distinct CP properties, i.e., $A\phi_a$. The
corresponding gauge structures can be assembled by combining the
Levi-Civita tensor and the 4-vectors $k_{1\mu}$, $k_{2\mu}$, and
$k_{3\mu}$. However, not all the combinations are independent as
some of them can be eliminated with the help of Schouten's
identity. Once those redundant structures are removed, we are left
with two independent gauge structures. We choose the following
basis
\begin{eqnarray}
&&\widetilde{P}_{1\mu \nu}=\frac{\sqrt{2}}{s}\epsilon_{\mu
\nu \alpha \beta}k^\alpha_1k^\beta_2, \\
&&\widetilde{P}_{2\mu
\nu}=\frac{2\sqrt{2}}{k^2_Ts}\Big[\frac{1}{2}k^2_3 \epsilon_{\mu
\nu \alpha \beta}k^\alpha_1 k^\beta_2+k_1\cdot k_3\epsilon_{\mu
\nu \alpha \beta}k^\alpha_2 k^\beta_3+k_{3\mu}\epsilon_{\nu \alpha
\beta \gamma}k^\alpha_1k^\beta_2 k^\gamma_3\Big],
\end{eqnarray}
which also are orthonormal.
The invariant amplitude can be written as follows:
\begin{equation}
{\cal M}={\cal M}_{\mu \nu}\epsilon^\mu
(k_1,\lambda_1)\epsilon^\nu (k_2,\lambda_2),
\end{equation}
where $\epsilon^\mu (k_1,\lambda_1)$ and $\epsilon^\nu
(k_2,\lambda_2)$ are the polarization vectors of the photons. The
tensor amplitude reads
\begin{equation}
{\cal M}_{\mu \nu}=\frac{\alpha^2}{\sqrt{2}s^2_W}\left\{
\begin{array}{lr}
A_1P_{1\mu \nu}+A_2P_{2\mu \nu}\quad {\rm for} & AA, \phi_a \phi_a, \\
\ \ \ \\
\widetilde{A}_1\widetilde{P}_{1\mu
\nu}+\widetilde{A}_2\widetilde{P}_{2\mu \nu}\quad {\rm for} &
A\phi_a
\end{array}\right.,
\end{equation}
and the unpolarized cross-section for the process $\gamma
\gamma \to \phi_i \phi_i$ is given by
\begin{eqnarray}
\sigma(\gamma \gamma \to \phi_i \phi_i)&=&\frac{1}{16\pi
s^2\epsilon}\int^{t_0}_{t_1}dt\sum_{spin}|{\cal
M}|^2 \nonumber\\
&& =\frac{\alpha^4}{128\pi
s^4_Ws^2\epsilon}\int^{t_0}_{t_1}dt\left\{
\begin{array}{ll}
(|A_1|^2+|A_2|^2), \\
\ \ \ \\
(|\widetilde{A}_1|^2+|\widetilde{A}_2|^2),
\end{array}\right.
\end{eqnarray}
where $\epsilon=2$ if the final particles are identical and 1
otherwise. The integration limits are
\begin{equation}
t_0(t_1)=-\frac{s}{2}\Big[1-\frac{k^2_3+k^2_4}{s}\mp
\sqrt{1-2\Big(\frac{k^2_3+k^2_4}{s}\Big)+\Big(\frac{k^2_3-k^2_4}{s}\Big)^2}\Big].
\end{equation}

\begin{figure}
\centering
\includegraphics[width=4in]{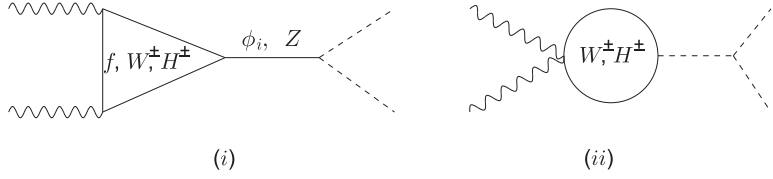}
\caption{\label{FIG1} Feynman diagrams contributing to the $\gamma
\gamma \to \phi_i \phi_j$ processes in the nonlinear gauge, in which
this contribution  is ultraviolet finite and gauge invariant by
itself. In the case of the $W$ contribution, similar diagrams are to
be considered for the pseudo-Goldstone and ghost fields. Crossed
diagrams are not shown. }
 \end{figure}

\begin{figure}
\centering
\includegraphics[width=4in]{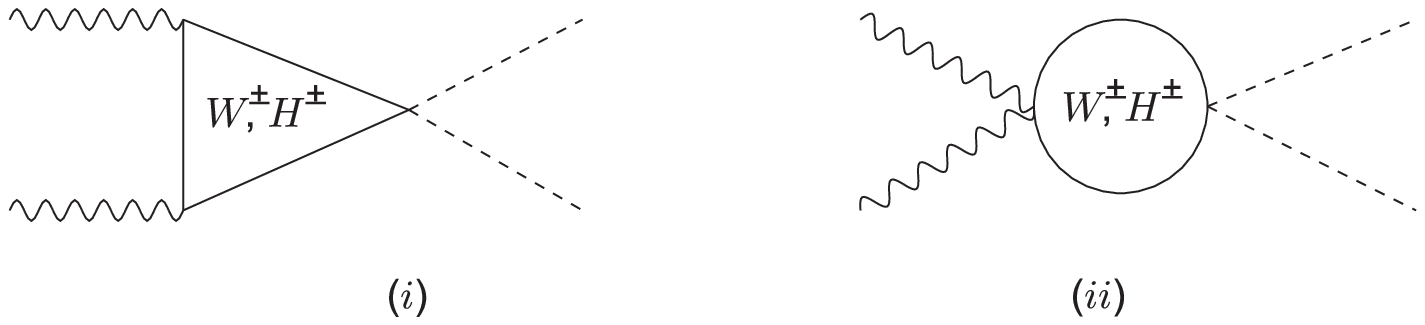}
\caption{\label{FIG2} The same as in Fig. \ref{FIG1}.}
\end{figure}

\begin{figure}
\centering
\includegraphics[width=4in]{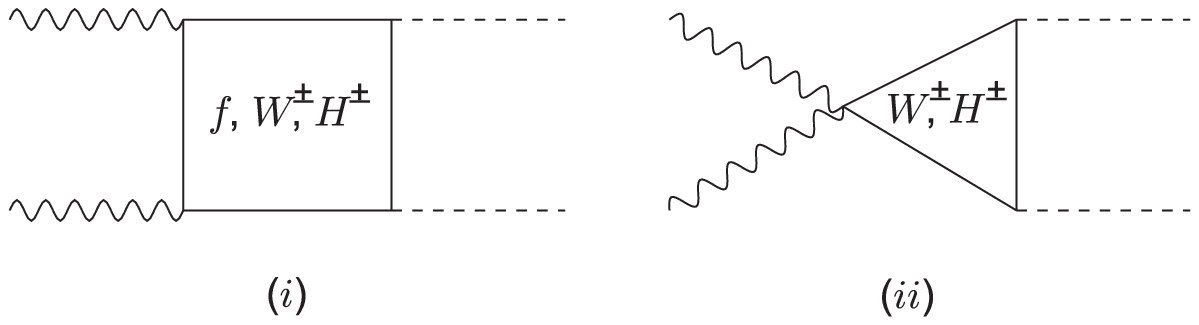}
\caption{\label{FIG3} The same as in Fig. \ref{FIG1}.}
\end{figure}

\begin{figure}
\centering
\includegraphics[width=4in]{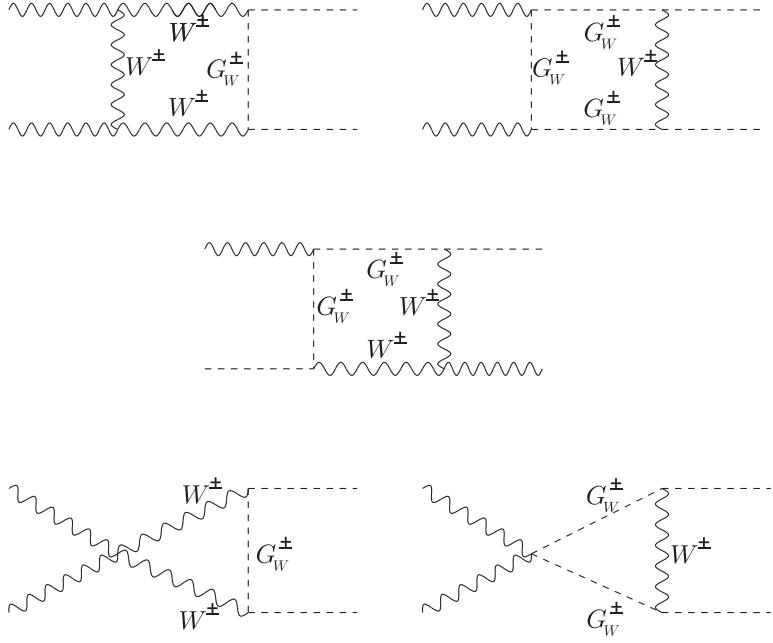}
\caption{\label{FIG4} Feynman diagrams contributing to the $\gamma
\gamma \to \phi_i \phi_j$ processes in the nonlinear gauge. The
contribution of these diagrams is finite and gauge invariant by
itself. Crossed diagrams are not shown.}
\end{figure}

\begin{figure}
\centering
\includegraphics[width=4in]{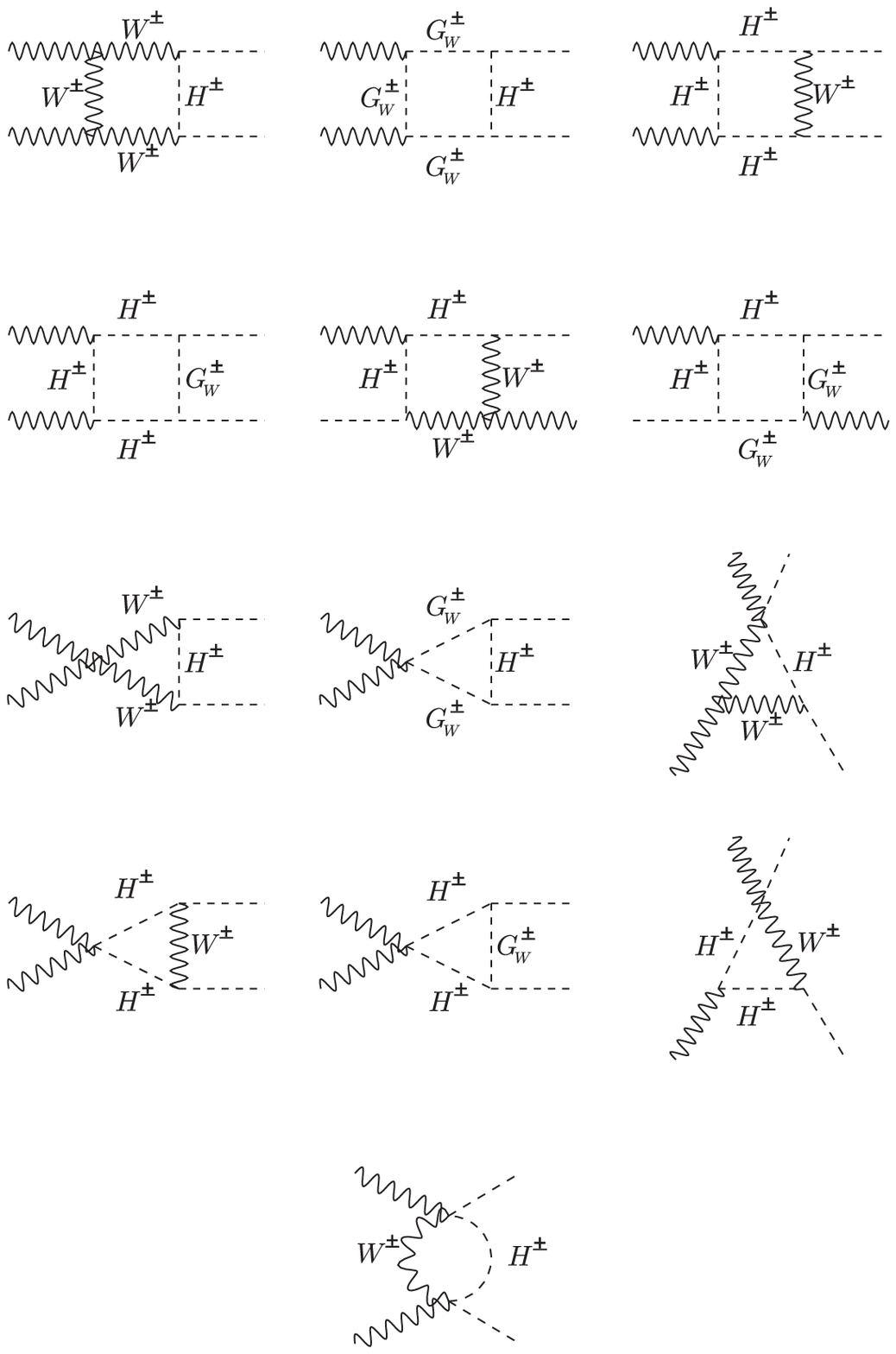}
\caption{\label{FIG5} The same as in Fig. \ref{FIG4}.}
\end{figure}

We will present below the amplitudes for the three available
processes: $\gamma \gamma \to AA$, $\gamma \gamma \to A\phi_a$, and
$\gamma \gamma \to \phi_a \phi_b$. The absence of the unphysical
$\gamma WG_W$ and $\phi_a\gamma WG_W$ vertices \footnote{The
$A\gamma WG_W$ vertex is not generated by the theory.} introduces
considerable simplifications in the calculations. In particular,
there are a significant reduction in the number of diagrams with
respect to those appearing in a linear gauge. Also, the
contributions can be grouped into distinct sets of diagrams which
lead to finite and gauge invariant results by their own (see Figs.
\ref{FIG1}-\ref{FIG5}).

We find it convenient to  express our results in terms of
Passarino-Veltman scalar functions. For those scalar functions
arising from the sets of diagrams of Figs. \ref{FIG1} through
\ref{FIG4}, an unambiguous shorthand notation can be used:
\begin{eqnarray}
C_0(a,b)&=&C_0(k^2_a,k^2_b,(k_a+k_b)^2,m^2,m^2,m^2), \\
D_0(a,b,c)&=&D_0(k^2_a,k^2_b,k^2_c,(k_a+k_b+k_c)^2,(k_a+k_b)^2,(k_b+k_c)^2,m^2,m^2,m^2,m^2),
\end{eqnarray}
where the right-hand side  has been expressed in the notation of
Ref. \cite{Mertig}. In addition, $a$, $b$, and $c$ run over 1 to 3,
and $m$ is the mass of the particle circulating in the loop. Below
it will be evident why it is not necessary to specify $m$ as an
argument of the scalar functions. Unfortunately this scheme cannot
be used for the scalar functions arising from Fig. \ref{FIG5} since
two different particles circulate in the loop, so an adequate
notation for these scalar functions will be given below.

It is also convenient to introduce the following dimensionless
variables $x_f=m^2_f/s$, $x_A=m^2_A/s$, $x_a=m^2_a/s$ ($m_a$
stands for the mass of $\phi_a$), $x_{H}=m^2_{H^\pm}/s$,
$x_W=m^2_W/s$, $x_T=k^2_T/s$, $x_t=t/s$, $x_u=u/s$, and
$\gamma_i=\Gamma_i/\sqrt{s}$, being $\Gamma_i$ the total width of
$\phi_i$.
\subsection{The process $\gamma \gamma \to AA$}
This process receives contributions from charged fermions, the $W$
gauge boson and the charged Higgs boson $H^\pm$. These
contributions arise through the diagrams shown in Figs.
\ref{FIG1}, \ref{FIG2}, \ref{FIG3}, and \ref{FIG5}. There are not
contributions from diagrams of Fig. \ref{FIG4} since the $AWG_W$
vertex is not generated by the theory. As pointed out before, in
the nonlinear gauge, each set of diagrams leads to an ultraviolet
finite and gauge invariant amplitude. All of the diagrams of Figs.
\ref{FIG1}, \ref{FIG2}, \ref{FIG3}, and \ref{FIG5} contribute to
the $A_1$ amplitude, but only the diagrams of Figs. \ref{FIG3} and
\ref{FIG5} contribute to $A_2$. These contributions can
conveniently written as:
\begin{eqnarray}
&&A_1=A_{11}+A_{12}+A_{13}+A_{15}, \\
&&A_2=A_{23}+A_{25},
\end{eqnarray}
where the subscript $j$ in $A_{ij}$ denotes the contributions coming
from each set of diagrams, which can be expressed in terms of the
contributions of each kind of particles circulating in the loops as
follows:
\begin{eqnarray}
&&A_{11}=F_{11}+S_{11}+G_{11},\\
&&A_{12}=S_{12}+G_{12},\\
&&A_{13}=F_{13}, \\
&&A_{23}=F_{23},
\end{eqnarray}
where we will use the letters $F_{ij}$, $S_{ij}$, and $G_{ij}$ to
denote the contributions coming from fermion, scalar, and gauge
particles, respectively. The set of diagrams in Fig. \ref{FIG1}
yield
\begin{eqnarray}
&&F_{11}=-\sum_{f}N_fQ^2_f\sum_{\phi_a}({\cal G}_{\phi_a\bar{f}f}{\cal G}_{\phi_a
AA})\frac{x_f}{1-x_a+i\sqrt{x_a}\gamma_a}[2+(4x_f-1)sC_0(1,2)],\\
&&S_{11}=-\sum_{\phi_a}\Big(\frac{{\cal G}_{\phi_aH^-H^+}{\cal G}_{\phi_aAA}}{4}\Big)\frac{x_W}{1-x_a+i\sqrt{x_a}\gamma_a}
[1+2x_{H}sC_0(1,2)],\\
&&G_{11}=-\sum_{\phi_a}\Big(\frac{{\cal G}_{\phi_aWW}{\cal
G}_{\phi_aAA}}{2}\Big)\frac{x_W}{1-x_a+i\sqrt{x_a}\gamma_a}
\Big[12+\frac{x_a}{x_W}+\Big(12x_W-4+x_a\Big)2sC_0(1,2)\Big],
\end{eqnarray}
where $N_f$ is the color index and $Q_f$ is the electric charge in
units of the positron charge, ${\cal G}_{\phi_a (ff,VV,\phi
\phi)}$ are functions of the couplings $g_{\phi_a (ff,VV,\phi
\phi)}$, which they are given in  Appendix. Notice that there is
no contribution from the one--loop $\gamma \gamma Z^*$ off--shell
coupling~\cite{Barroso} as the $Z$ boson only couples to pairs of
the form $A\phi_a$. In the set of diagrams of Fig. \ref{FIG2}
there are only contributions from the charged scalar boson and the
$W$ gauge boson. The corresponding amplitudes are
\begin{eqnarray}
&&S_{12}=-\Big(\frac{{\cal G}_{H^\pm H^\mp AA}}{4}\Big)[1+2x_{H}sC_0(1,2)], \\
&&G_{12}=-8\Big[1-(1-2x_W)sC_0(1,2)+\Big(\frac{{\cal
G}_{G^\pm_WG^\mp_WAA}}{32}\Big)[1+2x_WsC_0(1,2)]\Big].
\end{eqnarray}
There are contributions to $G_{12}$ from the $W$ boson and its
associated pseudo-Goldstone boson, but not from the ghost field
because of the absence of the $\bar{C}^\pm C^\mp AA$ vertex. As far
as the diagrams of Fig. \ref{FIG3} is concerned, there are no
contributions from scalar or gauge particles due to the absence of
the $H^\pm H^\mp A$ and $WWA$ vertices. The fermion contributions
are given by
\begin{eqnarray}
F_{13}&=&2\sum_fN_fQ^2_f{\cal
G}^2_{A\bar{f}f}\Big(\frac{x_f}{x_W}\Big)\Bigg\{2+4x_fsC_0(1,2)
+2x_As\Big[(x_t-x_A)C_0(2,3)+(x_u-x_A)C_0(1,3)\Big]+\nonumber
\\
&&s^2\Bigg[-x_Ax_TD_0(1,3,2)+x_f(1-2x_A)\Big[D_0(1,2,3)+D_0(1,3,2)+D_0(2,1,3)\Big]\Bigg]\Bigg\},
\end{eqnarray}
\begin{eqnarray}
F_{23}&=&2\sum_fN_fQ^2_f{\cal
G}^2_{A\bar{f}f}\Big(\frac{x_f}{x_W}\Big)\Bigg\{s\Bigg[C_0(1,2)+(2x_A-1)C_0(3,4)+(x_t-x_A)C_0(2,3)+(x_u-x_A)C_0(1,3)\nonumber
+\\
&&\frac{1-2x_A}{2x_T}\Big[2x_t(x_t-x_A)C_0(2,3)+2x_u(x_u-x_A)C_0(1,3)-(1-2x_A)C_0(1,2)-(x_t-x_u)^2C_0(3,4)\Big]\Bigg]+\nonumber
\\
&&\frac{s^2}{2x_T}\Bigg[-2(1-2x_A)x_fx_T\Big[D_0(1,2,3)+D_0(1,3,2)+D_0(2,1,3)\Big]+\nonumber
\\
&&x_t(x^2_t+x^2_A)D_0(1,2,3)+x_u(x^2_u+x^2_A)D_0(2,1,3)\Bigg]\Bigg\}.
\end{eqnarray}
Finally, the contributions from  Fig. \ref{FIG5} read
\begin{eqnarray}
&&A_{15}=2+s\sum^{12}_{a=3}f^1_aC_0(a)+s^2\sum^6_{a=1}g^1_aD_0(a),\\
&&A_{25}=\frac{1}{x_Wx_T}\Bigg[s\sum^{12}_{a=1}f^2_aC_0(a)+s^2\sum^6_{a=1}g^2_aD_0(a)\Bigg],
\end{eqnarray}
where
\begin{eqnarray}
&&f^1_3=-4x_{H}, \\
&&f^1_4=-8(1-x_W), \\
&&f^1_5=f^1_6=f^1_{11}=f^1_{12}=\frac{1}{x_W}
A_0(x_u-x_A),\\
&&f^1_7=f^1_8=f^1_9=f^1_{10}=\frac{1}{x_W}
A_0(x_t-x_A),\\
&&g^1_1=g^1_3=-2\Big(\frac{x_{H}}{x_W}\Big)
(2x_W+A_0),\\
&&g^1_2=g^1_4=-2
(2x_H+A_0),\\
&&g^1_5=g^1_6=-4x_H-\frac{(x_T+x_H+x_W)A_0}{x_W},
\end{eqnarray}
\begin{eqnarray}
f^2_1&=&f^2_2=(2x_W+A_0)\Big[2x_T+(x_t-x_u)^2\Big],
\\
f^2_3&=&(2x_W+A_0)(2x_W-2x_H-x_t-x_u)\\
f^2_4&=&(2x_W+A_0)(2x_H-2x_W-x_t-x_u)-8x_Tx_W; \\
f^2_5&=&f^2_6=f^2_{11}=f^2_{12}=(x_A-x_u)
\Big[x_u A_0-2x_W(x_A-x_u)^2\Big],\\
f^2_7&=&f^2_8=f^2_9=f^2_{10}=f^2_5(x_u\to x_t),
\end{eqnarray}
\begin{equation}
g^2_1=(A_0+2x_W)[2x_T x_H+(x_H+x_u-x_W)^2],
\end{equation}
\begin{eqnarray}
g^2_2&=&4x_W(x_A-x_u)^4-4x_W(x_A-x_u)^3
+(x_A-x_u)^2\Big[2(2x_A+2x_H-4x_W+1)x_W+A_0(1-2x_W)\Big]\nonumber\\
&&+2(x_A-x_u)\Big[2x_W(x_W-x_A)+A_0(x_H-x_A)\Big]+A_0^2+2x_W(x_A+1)A_0+4x_Ax_Hx_W,
\end{eqnarray}
\begin{eqnarray}
g^2_3&=&g^2_1(x_u \to x_t), \\
g^2_4&=&g^2_2(x_t \leftrightarrow x_u),
\end{eqnarray}
\begin{eqnarray}
g^2_5=g^2_6&=&\frac{1}{4}\Big\{(A_0+2x_W)(2x_H-2x_W+x_t+x_u)^2\nonumber\\
&&+2\big[A_0(x_T-x_t-x_u)-2x_W(x_t+x_u)\big](2x_H-2x_W+x_t+x_u)\nonumber\\
&&+8x_Tx_W(A_0+2x_W)+8x_T^2x_W-2A_0(x_t+x_u)x_T\nonumber\\
&&+(A_0+2x_W)(x_t+x_u)^2\Big\}.
\end{eqnarray}
with
\begin{equation}
A_0=x_A^2-2(x_H+x_W)x_A+(x_H-x_W)^2,
\end{equation}
The arguments for the $C_0 (a)$ and $D_0 (a)$ scalar functions are presented
in Appendix B.
\subsection{The process $\gamma \gamma \to \phi_a \phi_b$}
We turn now to the amplitudes associated with the final states $hh$,
$hH$, and $HH$. There are contributions from all the sets of
diagrams shown in Figs. \ref{FIG1} to \ref{FIG5}. According to our
notation, the contributions can be organized as follows:
\begin{eqnarray}
&&A_1=A_{11}+A_{12}+A_{13}+A_{14}+A_{15}, \\
&&A_2=A_{23}+A_{24}+A_{25},
\end{eqnarray}
where
\begin{eqnarray}
&&A_{11}=F_{11}+S_{11}+G_{11},\\
&&A_{12}=S_{12}+G_{12},\\
&&A_{13}=F_{13}+S_{13}+G_{13},
\end{eqnarray}
\begin{eqnarray}
A_{23}=F_{23}+S_{23}+G_{23}.
\end{eqnarray}
The $A_{14}$, $A_{15}$ $A_{24}$, and $A_{25}$ partial amplitudes
only receive contributions from the pairs $(W,G_W)$ and $(W,H^\pm)$.
The contributions coming from the set of diagrams in Fig. \ref{FIG1}
are given by
\begin{eqnarray}
&&F_{11}=\sum_fN_fQ^2_f\sum_{\phi_c}({\cal G}_{\phi_c\bar{f}f}{\cal G}_{\phi_c\phi_b\phi_a})\frac{x_f}{1-x_c+i\sqrt{x_c}\gamma_c}
[2+(4x_f-1)sC_0(1,2)],\\
&&S_{11}=-\sum_{\phi_c}\Big(\frac{{\cal G}_{\phi_cH^\pm
H^\mp}{\cal G}_{\phi_c\phi_b\phi_a}}{4}\Big)\frac{x_W}{1-x_c+i\sqrt{x_c}\gamma_c}[1+2x_HsC_0(1,2)],\\
&&G_{11}=-\sum_{\phi_c}\Big(\frac{{\cal G}_{\phi_cWW}{\cal
G}_{\phi_c\phi_b\phi_a}}{2}\Big)\frac{x_W}{1-x_c+i\sqrt{x_c}\gamma_c}
\Big[12+\frac{x_c}{x_W}+2(-4+12x_W+x_c)sC_0(1,2)\Big],
\end{eqnarray}
\begin{eqnarray}
&&S_{12}=\frac{-{\cal G}_{H^\pm H^\mp
\phi_a\phi_b}}{4}[1+2x_HsC_0(1,2)],
\\
&&G_{12}=8 sC_0(1,2)\delta^{ab}+(2C_0(1,2) s x_W+1)(2{\cal
G}_{\phi_aWW}{\cal G}_{\phi_bWW}-\frac{1}{4}{\cal G}_{G_W^\pm
G_W^\mp\phi_a\phi_b}-8\delta^{ab}),
\end{eqnarray}
\begin{eqnarray}
F_{13}&=&\sum_fN_fQ^2_f({\cal G}_{\phi_a\bar{f}f}{\cal
G}_{\phi_b\bar{f}f})\Big(\frac{x_f}{x_W}\Big)\Bigg\{4\Big[1+2x_fsC_0(1,2)\Big]+\nonumber
\\
&&(1+x_t+x_u-8x_f)s\Big[(x_u-x_a)C_0(1,3)+(x_u-x_b)C_0(2,4)+(x_t-x_a)C_0(2,3)+(x_t-x_b)C_0(1,4)\Big]-\nonumber
\\
&&2x_f(1+x_a+x_b-8x_f)s^2\Big[D_0(1,2,3)+D_0(2,1,3)\Big]+\nonumber
\\
&&\Big[16x^2_f+2x_f(4x_T-x_a-x_b-1)-x_T(x_a+x_b)\Big]s^2D_0(1,3,2)\Bigg\},
\end{eqnarray}
\begin{eqnarray}
S_{13}&=&-\Big(\frac{{\cal G}_{\phi_a H^\pm H^\mp }{\cal
G}_{\phi_b H^\pm H^\mp
}}{4}\Big)x_Ws\Bigg\{(x_a-x_u)C_0(1,3)+(x_b-x_u)C_0(2,4)+(x_a-x_t)C_0(2,3)+\nonumber
\\
&&(x_b-x_t)C_0(1,4)+2x_Hs\Big[D_0(1,2,3)+D_0(2,1,3)+D_0(1,3,2)\Big]+x_TsD_0(1,3,2)\Bigg\},
\end{eqnarray}
\begin{eqnarray}
G_{13}&=&\Big(-\frac{{\cal G}_{\phi_aWW}{\cal G}_{\phi_bWW}}{x_W}\Big)s\Bigg\{A_G\Big[(x_a-x_u)C_0(1,3)+(x_b-x_u)C_0(2,4)+(x_a-x_t)C_0(2,3)\nonumber\\
&&+(x_b-x_t)C_0(1,4)\Big]+2x_W s(A_G-4x_W)\Big[D_0(1,2,3)+D_0(2,1,3)+D_0(1,3,2)\Big]+ \nonumber \\
&& s x_TA_GD_0(1,3,2)\Bigg\},
\end{eqnarray}
with
\begin{equation}
A_G=16x_W^2+x_bx_a+2x_W(x_a+x_b).
\end{equation}
\begin{eqnarray}
A_{14}&=&(4{\cal G}_{\phi_a W W}{\cal G}_{\phi_bW
W})s\Bigg\{-2C_0(1,2)+\nonumber
\\
&&(x_a+x_b)\Big[(x_a-x_u)C_0(1,3)+(x_b-x_u)C_0(2,4)+(x_a-x_t)C_0(2,3)+(x_b-x_t)C_0(1,4)\Big]+\nonumber
\\
&&2x_W(x_t+x_u)s\Big[D_0(1,2,3)+D_0(2,1,3)+D_0(1,3,2)\Big]+x_T(x_a+x_b)sD_0(1,3,2)\Bigg\},
\end{eqnarray}
\begin{equation}
A_{15}=\Big({\cal G}_{W^\pm H^\mp \phi_a}{\cal G}_{W^\pm H^\mp
\phi_b}\Big)\Big[2+s\sum^{12}_{a=
3}f^1_aC_0(a)+s^2\sum^6_{a=1}g^1_aD_0(a)\Big],
\end{equation}
where
\begin{eqnarray}
&&f^1_3=-4x_H, \\
&&f^1_4=8(x_W-1), \\
&&f^1_5=f^1_6=-(x_a-x_u)f,\\
&&f^1_7=f^1_8=-(x_a-x_t)f,\\
&&f^1_9=f^1_{10}=(x_b-x_t)f, \\
&&f^1_{11}=f^1_{12}=(x_b-x_u)f,
\end{eqnarray}
\begin{eqnarray}
&&g^1_1=g^1_3=-2x_H(f+2), \\
&&g^1_2=g^1_4=-2x_W\Big(f+2\frac{x_H}{x_W}\Big), \\
&&g^1_5=g^1_6=-\Big(x_H+x_W+x_T \Big)f - 4x_Hx_W,
\end{eqnarray}
with
\begin{equation}
f=\frac{(x_H-x_W)^2-(x_H+x_W)(x_a+x_b)+x_ax_b}{x_W}.
\end{equation}
\begin{eqnarray}
F_{23}&=&\sum_fN_fQ^2_f({\cal G}_{\phi_a\bar{f}f}{\cal
}\Big)s\Bigg\{
-\Big[\frac{x^2_t+x^2_u-8x_f(x_t+x_u)+2x_ax_b}{x_T}\Big]C_0(1,2)+\nonumber
\\
&&\Big(\frac{x^2_t+x^2_u-2x_ax_b}{x_T}\Big)(x_t+x_u-8x_f)C_0(3,4)+\nonumber
\\
&&\Big[\frac{x_t(x_t-8x_f)+x_ax_b}{x_T}\Big]\Big[(x_a-x_t)C_0(2,3)+(x_b-x_t)C_0(1,4)\Big]+\nonumber
\\
&&\Big[\frac{x_u(x_u-8x_f)+x_ax_b}{x_T}\Big]\Big[(x_a-x_u)C_0(1,3)+(x_b-x_u)C_0(2,4)\Big]+\nonumber
\\
&&2x_f(x_t+x_u-8x_f)sD_0(1,3,2)+\nonumber \\
&&s\Big[-16x^2_f+2x_f(x_t+x_u-4\frac{x_t^{2}}{x_T})+(x^2_t+x_ax_b)\frac{x_t}{x_T}\Big]D_0(1,2,3)+\nonumber
\\
&&s\Big[-16x^2_f+2x_f(x_t+x_u-4\frac{x_u^{2}}{x_T})+(x^2_u+x_ax_b)\frac{x_u}{x_T}\Big]D_0(2,1,3)\Bigg\},
\end{eqnarray}
\begin{eqnarray}
S_{23}&=&\Big(\frac{{\cal G}_{\phi_aH^\pm H^\mp }{\cal
G}_{\phi_bH^\pm H^\mp
}}{4}\Big)\Big(\frac{x_W}{x_{T}}\Big)s\Big(A+x_HB\Big)\\
G_{23}&=&({\cal G}_{\phi_aWW}{\cal
G}_{\phi_bWW})\Big(\frac{1}{x_Tx_W}\Big)A_Gs\Big(A+x_WB \Big),
\end{eqnarray}
with $A$ and $B$  given by
\begin{eqnarray}
A&=&-(x_t+x_u)C_0(1,2)+(x^2_t+x^2_u-2x_ax_b)C_0(3,4)+x_t(x_a-x_t)C_0(2,3)+x_t(x_b-x_t)C_0(1,4)+\nonumber
\\
&&x_u(x_a-x_u)C_0(1,3)+x_u(x_b-x_u)C_0(2,4)+s\Big[x^2_tD_0(1,2,3)+x^2_uD_0(2,1,3)\Big],\\
B&=&2x_Ts\Big[D_0(1,2,3)+D_0(2,1,3)+D_0(1,3,2)\Big].
\end{eqnarray}
Finally, the $A_{24}$ and $A_{25}$ amplitudes read
\begin{eqnarray}
A_{24}&=&(4{\cal G}_{\phi_aW W}{\cal G}_{\phi_bW
W})\Big(\frac{s}{x_T}\Big)\Bigg\{(x^2_t+x^2_u+2x_ax_b)C_0(1,2)-(x^2_t+x^2_u-2x_ax_b)(x_t+x_u)C_0(3,4)-\nonumber
\\
&&(x^2_t+x_ax_b)\Big[(x_a-x_t)C_0(2,3)+(x_b-x_t)C_0(1,4)\Big]
-(x^2_u+x_ax_b)\Big[(x_a-x_u)C_0(1,3)+(x_b-x_u)C_0(2,4)\Big]+\nonumber
\\
&&s\Bigg(x_T\Big[-2x_W(x_t+x_u)+x_T\Big]D_0(1,3,2)+\Big[-2x_Wx_T(x_u+x_t)+x_T^2-x_t(x_ax_b+x_t^2)\Big]D_0(1,2,3)+\nonumber
\\
&&\Big[-2x_Wx_T(x_u+x_t)+x_T^2-x_u(x_ax_b+x_u^2)\Big]D_0(2,1,3)\Bigg)\Bigg\},
\end{eqnarray}
\begin{eqnarray}
A_{25}&=&\Big({\cal G}_{W^\pm H^\mp\phi_a}{\cal G}_{W^\pm H^\mp
\phi_b}\Big)\Big(\frac{s}{x_T}\Big)
\Bigg[\sum^{12}_{a=1}f^2_aC_0(a)+s\sum^6_{a=1}g^2_aD_0(a)\Bigg],
\end{eqnarray}
where
\begin{eqnarray}
f^2_1&=&f^2_2=(f+2)\big[(x_t-x_u)^2+2x_T\big]\\
f^2_3&=&-(f+2)(2x_H-2x_W+x_u+x_t),\\
f^2_4&=&(f+2)(2x_H-2x_W-x_u-x_t)-8x_T\\
f^2_5&=&f^2_6=(x_a-x_u)\big[fx_u-2(x_a-x_u)(x_b-x_u)\big],\\
f^2_7&=&f^2_8=(x_a-x_t)\big[fx_t-2(x_a-x_t)(x_b-x_t)\big],\\
f^2_9&=&f^2_{10}=(x_b-x_t)\big[fx_t-2(x_a-x_t)(x_b-x_t)\big],\\
f^2_{11}&=&f^2_{12}=(x_b-x_u)\big[fx_u-2(x_a-x_u)(x_b-x_u)\big],
\end{eqnarray}
On the other hand,
\begin{eqnarray}
g^2_1&=&(f+2)\Big[2x_Tx_H+(x_H-x_W+x_u)^2\Big],\\
g^2_2&=&\frac{1}{4}\Big\{(f+2)(2x_H-2x_W+x_t+x_u)^2
-2\Big[4x_T+(f+2)(x_t+3x_u)\Big](2x_H-2x_W+x_t+x_u)\nonumber\\
&&+f\Big[8x_Tx_W+(x_t+3x_u)^2\Big]+2\Big[8x_T^2+4(2x_W+x_t+3x_u)x_T+(x_t+3x_u)^2
\Big]\Big\},\\
g^2_3&=&(f+2)\Big[2x_Tx_H+(x_H-x_W+x_t)^2\Big],\\
g^2_4&=&\frac{1}{4}\Big\{(f+2)(2x_H-2x_W+x_t+x_u)^2
-2\Big[4x_T+(f+2)(x_u+3x_t)\Big](2x_H-2x_W+x_t+x_u)\nonumber\\
&&+f\Big[8x_Tx_W+(x_u+3x_t)^2\Big]+2\Big[8x_T^2+4(2x_W+x_u+3x_t)x_T+(x_u+3x_t)^2
\Big]\Big\},\\
g^2_5=g^2_6&=&\frac{1}{4}\Big\{(f+2)(2x_H-2x_W+x_t+x_u)^2+2\Big[f(x_T-x_t-x_u)-2(x_t+x_u)\Big]
(2x_H-2x_W+x_t+x_u)\nonumber\\
&&+8x_Tx_W(f+2)+8x_T^2-2f(x_t+x_u)x_T+(f+2)(x_t+x_u)^2\Big\}.
\end{eqnarray}
\subsection{The process $\gamma \gamma \to A\phi_a$}
Bosonic loops do not contribute to this process. The fermionic
contributions are given through diagrams of the type $(i)$ in Figs.
\ref{FIG1} and \ref{FIG3}. The corresponding amplitudes can be
written as follows:
\begin{eqnarray}
&&\widetilde{A}_1=\widetilde{F}_{11}+\widetilde{F}_{13}, \\
&&\widetilde{A}_2=\widetilde{F}_{23},
\end{eqnarray}
where, the subscript $j$  in $\widetilde{F}_{ij}$ stands for the
contribution of the particular set of diagrams. Using the same
notation defined above, the partial amplitudes read
\begin{equation}
\widetilde{F}_{11}=\widetilde{F}^A_{11}+\widetilde{F}_{11}^Z,
\end{equation}
where $\widetilde {F}_{11}^A$ ($\widetilde {F}_{11}^Z$) represents
the contribution due to Higgs boson $A$ ($Z$ boson).
We can write these factors as
\begin{eqnarray}
\widetilde{F}_{11}^A&=&-\sum_fN_fQ^2_f\Big({\cal
G}_{A\bar{f}f}{\cal G}_{\phi_aAA}\Big)
\frac{x_f}{1-x_A+i\sqrt{x_A}\gamma_A}sC_0(1,2),\\
\label{rb}
\widetilde{F}_{11}^Z&=&\frac{1}{32\pi^3 c_W^2}
\frac{x_u-x_t}{1-x_Z+i\sqrt{x_Z}\gamma_Z}{\cal G}_{ZA\phi_a}\sum_fg_A^fQ_f^2\int\limits_0^1dx_1\int\limits_0^{1-x_1}dx_2\frac{x_1x_2}{x_f-x_1x_2}.
\end{eqnarray}
The expression given in Eq.(\ref{rb}) corresponds to the one given in Ref.~\cite{Barroso}.
\begin{eqnarray}
\widetilde{F}_{13}&=&\sum_fN_fQ^2_f\Big({\cal G}_{A\bar{f}f}{\cal
G}_{\phi_a\bar{f}f}\Big)\Big(\frac{x_f}{x_W}\Big)  s
\Big\{x_T(x_A-x_a)sD_0(1,3,2)+\nonumber
\\
&&2x_f(1+x_A-x_a)s\Big[D_0(1,2,3)+D_0(2,1,3)+D_0(1,3,2)\Big]+\nonumber
\\
&&(x_a-x_A)\Big[(x_t-x_a)C_0(2,3)+(x_t-x_A)C_0(1,4)+(x_u-x_a)C_0(1,3)+(x_u-x_A)C_0(2,4)\Big]\Big\},
\end{eqnarray}
\begin{eqnarray}
\widetilde{F}_{23}&=&\sum_fN_fQ^2_f\Big({\cal G}_{A\bar{f}f}{\cal
G}_{\phi_a\bar{f}f}\Big)\Big(\frac{x_f}{x_W}\Big) s
\Big\{s\Big[x_t\Big(\frac{x_ax_A-x^2_t}{x_T}\Big)D_0(1,2,3)-x_u\Big(\frac{x_ax_A-x^2_u}{x_T}\Big)D_0(2,1,3)\Big]+\nonumber
\\
&&2x_f(x_u-x_t)s\Big[D_0(1,2,3)+D_0(2,1,3)+D_0(1,3,2)\Big]+\nonumber
\\
&&\Big(\frac{x_ax_A-x^2_u}{x_T}\Big)\Big[(x_u-x_a)C_0(1,3)+(x_u-x_A)C_0(2,4)\Big]-\nonumber
\\
&&\Big(\frac{x_ax_A-x^2_t}{x_T}\Big)\Big[(x_t-x_a)C_0(2,3)+(x_t-x_A)C_0(1,4)\Big]-\nonumber
\\
&&(x_u-x_t)\Big[\Big(\frac{x_t+x_u}{x_T}\Big)C_0(1,2)+\Big(\frac{4x_ax_A-(x_t+x_u)^2}{x_T}\Big)C_0(3,4)\Big]\Big\}.
\end{eqnarray}
\section{Numerical results and discussion}
\label{discussion}
  We now turn to discuss our results. We refer  to our three benchmark scenarios already discussed, namely,  (1)
 Scenario I (the decoupling limit), (2) Scenario II ( SM-like), and (3) Scenario III (a more general case of 2HDM-III).

\subsection{Scenario I (the decoupling limit)}
The main purposes of this subsection is to show explicitly how the decoupling of the charged Higgs effects operate according the criteria established in Ref.~\cite{Gunion}. In this scenario, one assumes the existence of a light Higgs boson $h$, with mass of order of the Fermi scale $v$, whereas the rest of Higgs bosons are assumed very heavy, \textit{i.e.}, $m_{H^\pm}\sim m_{A}\sim m_{H}>> v$. The heavy Higgs boson effects decouple through two essentially different mechanisms~\cite{Gunion}, namely, by assuming $\mu_{12}^2>>v^2$ or $\mu_{12}^2\sim v^2$ but taking large $\tan\beta$ or $\cot\beta$, depending on the configuration chosen for the $\lambda_6$ and $\lambda_7$ parameters. Accordingly, we will consider the following two cases:
\begin{description}
    \item[Case A] $m_h=120\ GeV$, $\alpha=\beta-\pi/2$, $\lambda_6=\lambda_7=0$,
    $\mu_{12}^2>>v^2$, and $m_H^+>>m_h$.
    \item[Case B] $m_h=120\ GeV$, $\alpha=\beta-\pi/2$, $\lambda_6=-\lambda_7=0.1$,
    $\mu_{12}^2\sim v^2$, and $m_H^+>>m_h$.
\end{description}
\begin{figure}[h]
\centering
\includegraphics[width=5 in]{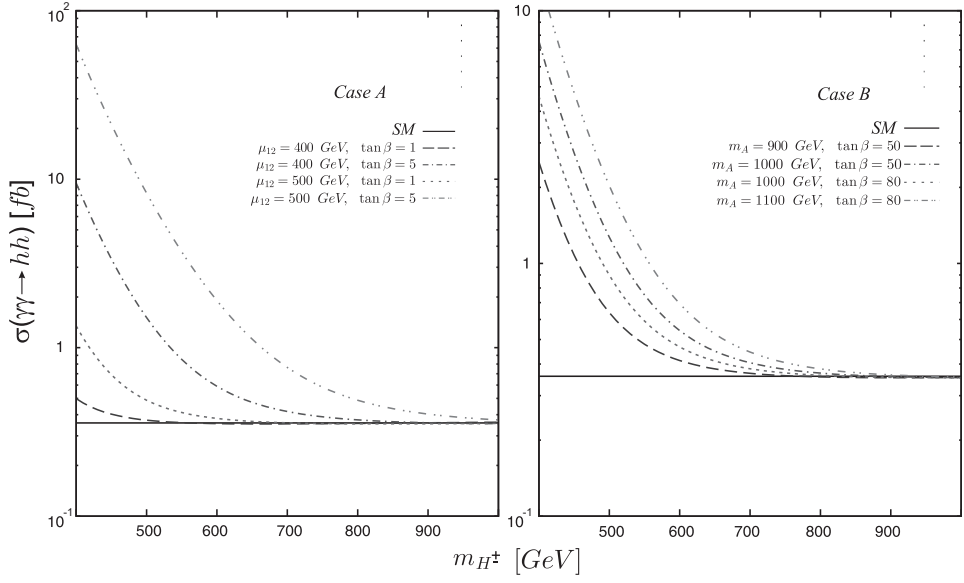}
\caption{Behavior of cross section for the process
$\gamma\gamma\to hh$ as a function of  the charged
Higgs mass in  the Scenario I (decoupling limit). The diverse values of the parameters as well as the case in consideration (A or B) are shown in the figure. In all cases the center-of-mass energy $\sqrt{s}=500$ GeV was used.  }
\label{dl-1}
\end{figure}
In the Figure \ref{dl-1}, we present the cross section of the process $ \sigma(\gamma \gamma \to hh)$  as a function  of the charged Higgs mass  $m_{H^\pm}$. One can get the SM result (represented in these figures by the horizontal thick line)  when $m_{H^\pm} >>m_h$, showing the decoupling nature of this contribution. On the  other hand, we can see from Fig. \ref{dl-2} that in the limit $\tan \beta <<1 $ of the case A the  SM result is obtained. The decoupling nature of the contributions for the case B is clearly appreciated when $\tan \beta >> 1$.

\begin{figure}[h]
\centering
\includegraphics[width=5 in]{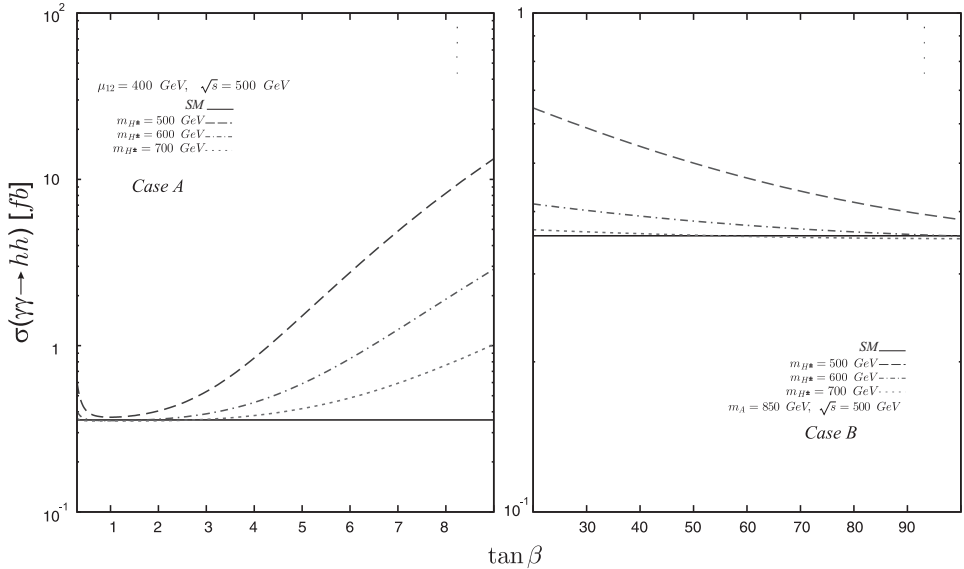}
\caption{Behavior of the cross section for the process
$\gamma\gamma\to hh$ as a function of  $\tan \beta$ in the Scenario I (decoupling limit). The diverse values of the parameters as well as the case in consideration (A or B) are shown in the figure. In all cases the center-of-mass energy $\sqrt{s}=500$ GeV was used.}
\label{dl-2}
\end{figure}

\subsection{Scenario II ( SM-like)}
As already commented, in this scenario the couplings $hVV$ ($V=W,Z$), $hhh$, $hhhh$ are nearly indistinguishable from the corresponding ones of the SM, but the $hf\bar{f}$ couplings can deviate significantly from their SM counterparts $h_{SM} f \bar{f}$. In this context, we will use the values $m_h=120\ GeV$, $m_A= 110$ GeV, $m_H=m_{H^\pm}=m_A+m_h$, and $\mu_{12}=130$ GeV in the two cases $(\lambda_6=-\lambda_7=0.1)$ and $(\lambda_6=-\lambda_7=1)$. The values $\chi_{uu,dd}=1$, $-1$, which arise from a selection of a specific texture of the Yukawa matrices, will be used. In addition, we will assume that $\alpha=\beta \pm \pi /2$.
 \begin{figure}[h]
\centering
\includegraphics[width=5 in]{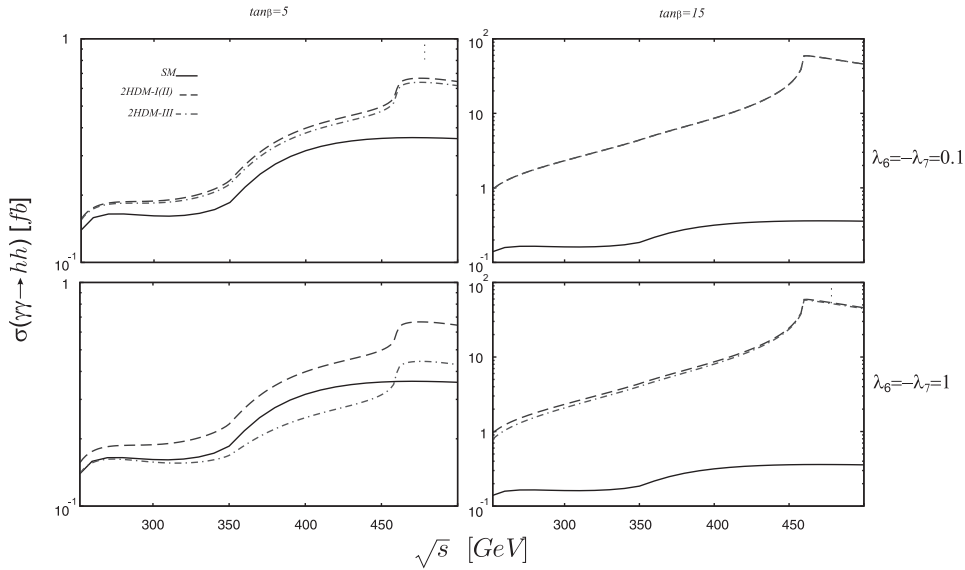}
\caption{Behavior of cross section for the process
$\gamma\gamma\to hh$ as a function of  center-of-mass energy $\sqrt{s}$ in the Scenario II (SM-like).  The curves correspond to
SM (solid line), 2HDM type I-II (dashed line),  and 2HDM-III (dashed-dotted line), for  $\tan\beta =5$ (left), 15 (right) in the case $\lambda_6=-\lambda_7=0.1$ (up panels), 1 (down panels).  }
\label{sml-1}
\end{figure}

In Fig. \ref{sml-1}, the cross section for the process $ \gamma \gamma \to hh$ as a function of center-of-mass energy $\sqrt{s}$ is shown. It can be appreciated that for $\tan \beta =5$ both the SM and the 2HDM's predictions essentially coincide. However, one can see that for $\tan \beta= 15$ the cross section predicted by the 2HDM«s could be two order of magnitude larger than the SM result. The results for $\chi=1, \, -1$ are very similar. In this case, the cross section of the mode $ \sigma(\gamma \gamma \to hh)\sim 70 $ fb.

\begin{figure}[h]
\centering
\includegraphics[width=5 in]{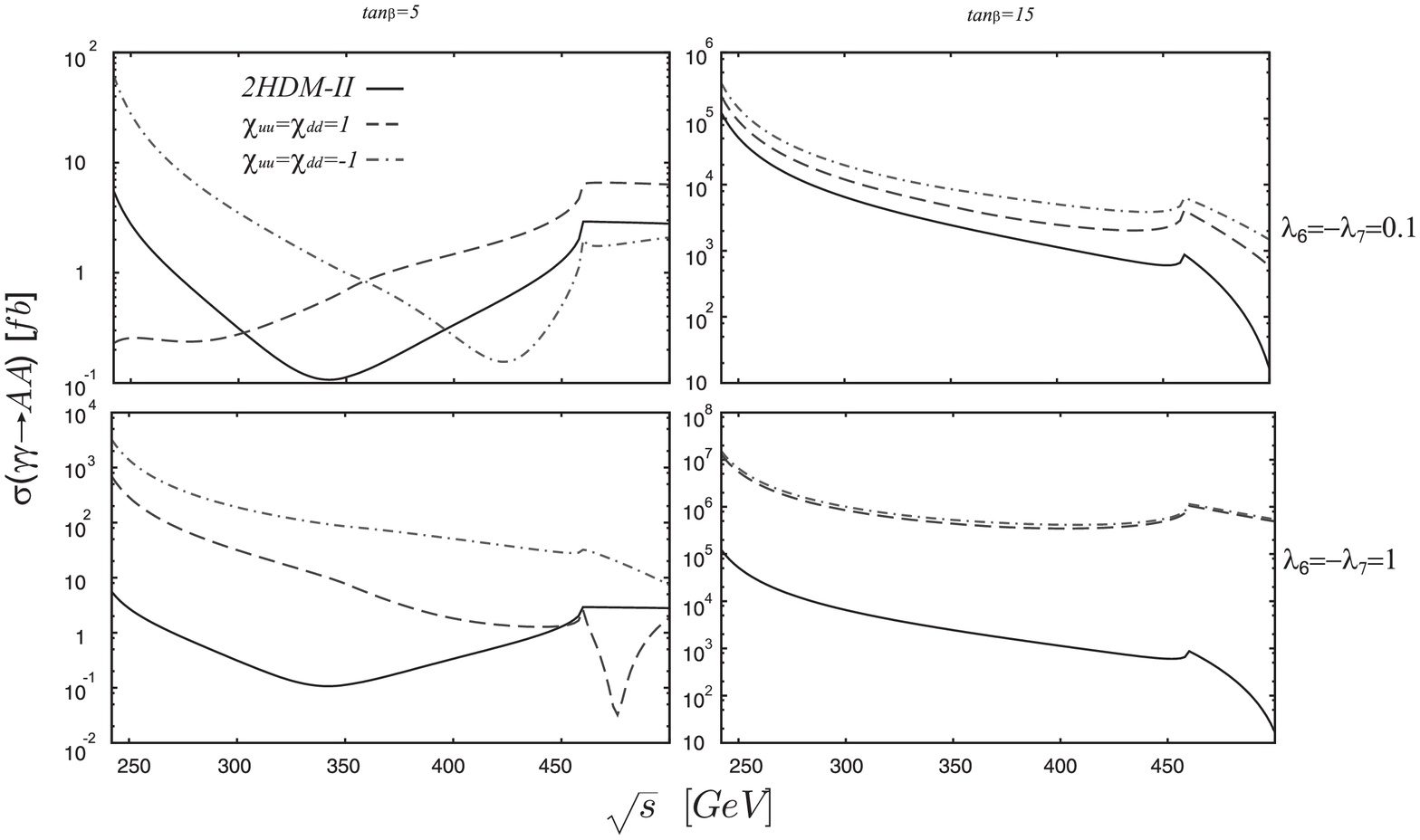}
\caption{Behavior of the cross section for the process
$\gamma\gamma\to AA$ as a function of  the center-of-mass
energy $\sqrt{s}$ in the Scenario II (SM-like).  The lines correspond to
2HDM-II (solid line), 2HDM-III with $\chi_{uu,dd}=1$ (dashed line)  and 2HDM-III when $\chi_{uu,dd}=-1$ (dashed-dotted line), for  $\tan\beta =5$ (left), 15  (right) in the cases $\lambda_6=-\lambda_7=0.1$ (up panels) and $\lambda_6=-\lambda_7=1$ (down panels). }
\label{sml-2}
\end{figure}

\begin{figure}[h]
\centering
\includegraphics[width=5 in]{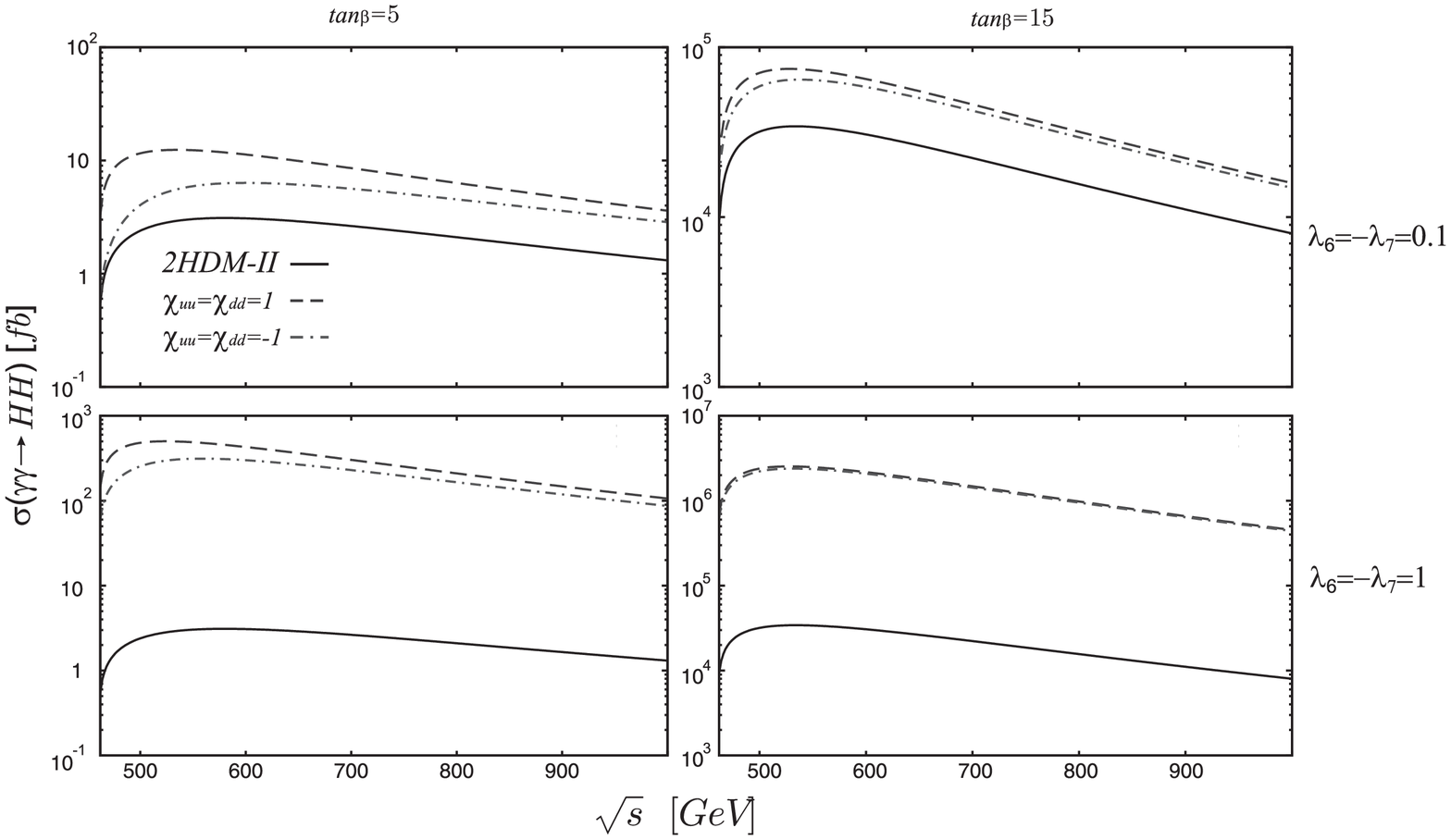}
\caption{Behavior of the cross section for the process
$\gamma\gamma\to HH$ as a function of  the center-of-mass
energy $\sqrt{s}$ in the Scenario II (SM-like). The description of the plots is the same as in Fig. \ref{sml-2}.}
\label{sml-3}
\end{figure}

On the other hand, in Fig. \ref{sml-2}, the cross sections for the $ \gamma \gamma \to AA$ reaction as a function of the center-of-mass
energy $\sqrt{s}$ is displayed and the predictions of the 2HDM-III compared with those generated by the 2HDM-II. In this process we also consider the contribution of the parameters of the Higgs potential $(\lambda_6=-\lambda_7=0.1)$ (up panels) and $(\lambda_6=-\lambda_7=1)$ (down panels). We can observe that the main impact for the cross section comes from $(\lambda_6=-\lambda_7=1)$ and $\tan \beta$ large.  The cross section of this mode could be enhanced by two orders of magnitude compared with the case $\lambda_6=-\lambda_7=0.1$ or the usual case $\lambda_6=-\lambda_7=0$ (2HDM-II). One can get
$ \sigma(\gamma \gamma \to AA)\sim 5 \times 10^5$ fb for $\tan \beta = 15$, with $\sqrt{s}= 450$ GeV, taking $\chi=1, \, -1$  and  $\lambda_6=-\lambda_7=1$. The results for the cross section of the process $ \gamma \gamma \to HH$  are shown in Fig. \ref{sml-3} for the same parameters of the previous process.  Likewise, the cross section predicted by the 2HDM-III is two orders of magnitude larger than the one predictied by the 2HDM-II. We can obtain $ \sigma(\gamma \gamma \to HH)\sim 1 \times 10^6$ fb for $\tan \beta = 15$, $\chi=1, \, -1$, with $\sqrt{s}= 500$ GeV and  $\lambda_6=-\lambda_7=1$.
\begin{figure}[h]
\centering
\includegraphics[width=5 in]{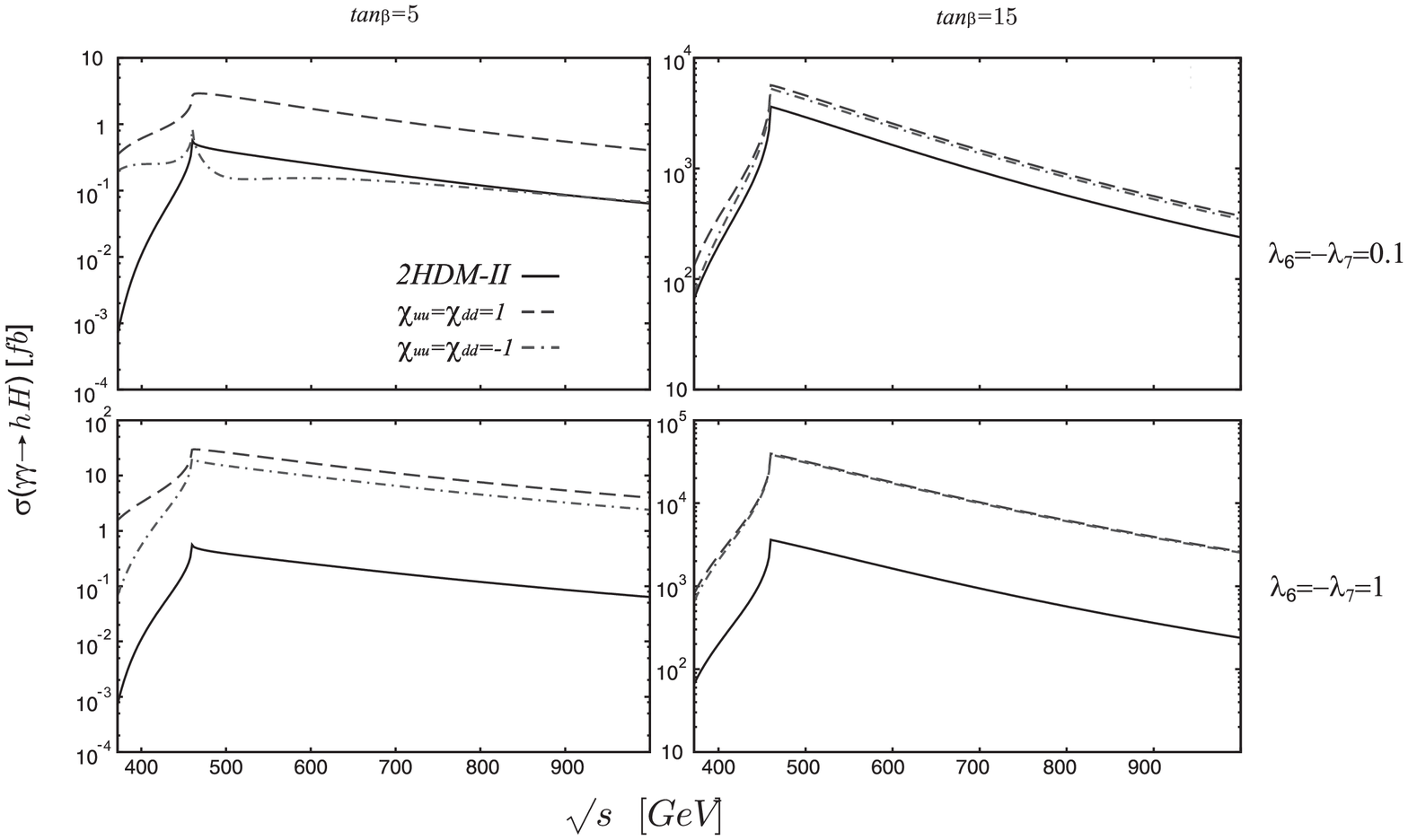}
\caption{Behavior of the cross section for the process
$\gamma\gamma\to hH$ as a function of  the center-of-mass
energy $\sqrt{s}$ in the Scenario II (SM-like). The description of the plots is the same as in Fig. \ref{sml-2}. }
\label{sml-4}
\end{figure}
As far as the cross section for the $ \sigma(\gamma \gamma \to hH)$ process is concerned, its behavior as a function of the center-mass energy $\sqrt{s}$ is shown in Fig. \ref{sml-4}. From this figure, a considerable enhancement of the cross section compared with the 2HDM-II prediction can be observed for $\lambda_6=1=-\lambda_7$. In this case, $ \sigma(\gamma \gamma \to hH)\sim 5 \times 10^4$ fb for $\tan \beta = 15$, around $\sqrt{s}= 470$ GeV and $\chi=1, \, -1$ . It is worth commenting that this is the first time that this process is studied. Another process which has not been studied in the literature is $ \sigma(\gamma \gamma \to hA)$. The corresponding cross sections as a function of the center-mass energy $\sqrt{s}$ is shown in Fig. \ref{sml-5}. It can be appreciated from this figure the importance of the cross section for the case $\lambda_6=1=-\lambda_7$ and $\tan \beta = 15$, which is quite large as compared with the prediction of 2HDM-II where the $\lambda_6$ and $\lambda_7$ parameters are absent. It can be appreciated that the cross section could be of the order of 7 fb for $\sqrt{s} = 350$ GeV. On the other hand, when  $\lambda_6=-\lambda_7 << 1$ the cross section is very insignificant to be considered as relevant signals of neutral Higgs bosons. The last numerical results of this scenario is the cross section for the process
$\gamma\gamma\to HA$, which is shown as a function of the center-mass energy in Fig. \ref{sml-6}. It can appreciated an important value for the cross section of about $2 \times 10^3 \, fb$ for the case $\lambda_6=-\lambda_7=1$, $\chi =  -1$,  $\tan \beta = 15$, and $\sqrt{s}$ around $350$ GeV.
\begin{figure}[h]
\centering
\includegraphics[width=4 in]{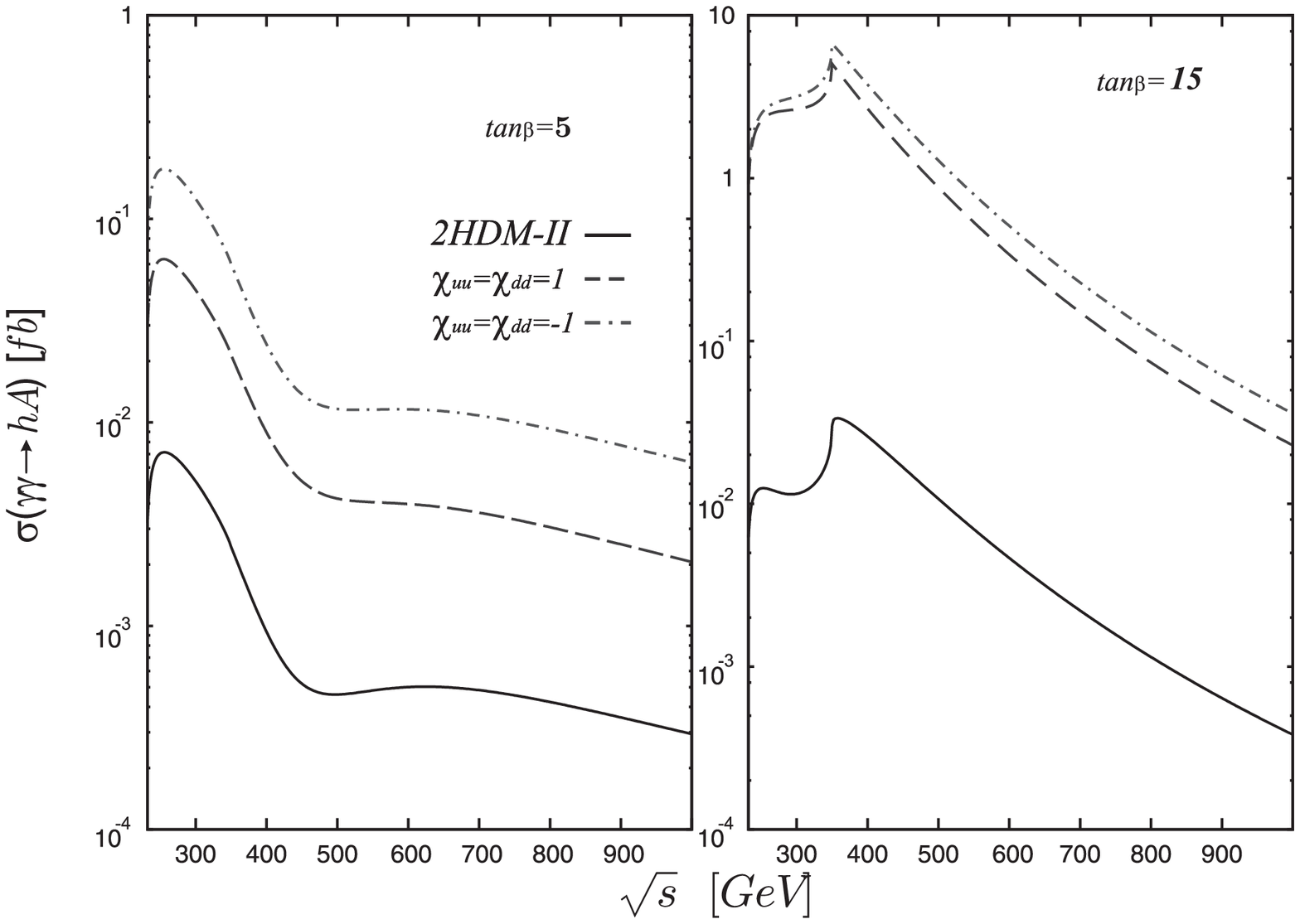}
\caption{Behavior of the cross section for the process
$\gamma\gamma\to hA$ as a function of  the center-of-mass
energy $\sqrt{s}$ in the Scenario II (SM-like). The lines correspond to
2HDM-II (solid line), 2HDM-III with $\chi_{uu,dd}=1$ (dashed line)  and 2HDM-III when $\chi_{uu,dd}=-1$ (dashed-dotted line), for  $\tan\beta =5$ (left), 15  (right) in the case $\lambda_6=-\lambda_7=1$.  }
\label{sml-5}
\end{figure}
\begin{figure}[h]
\centering
\includegraphics[width=5 in]{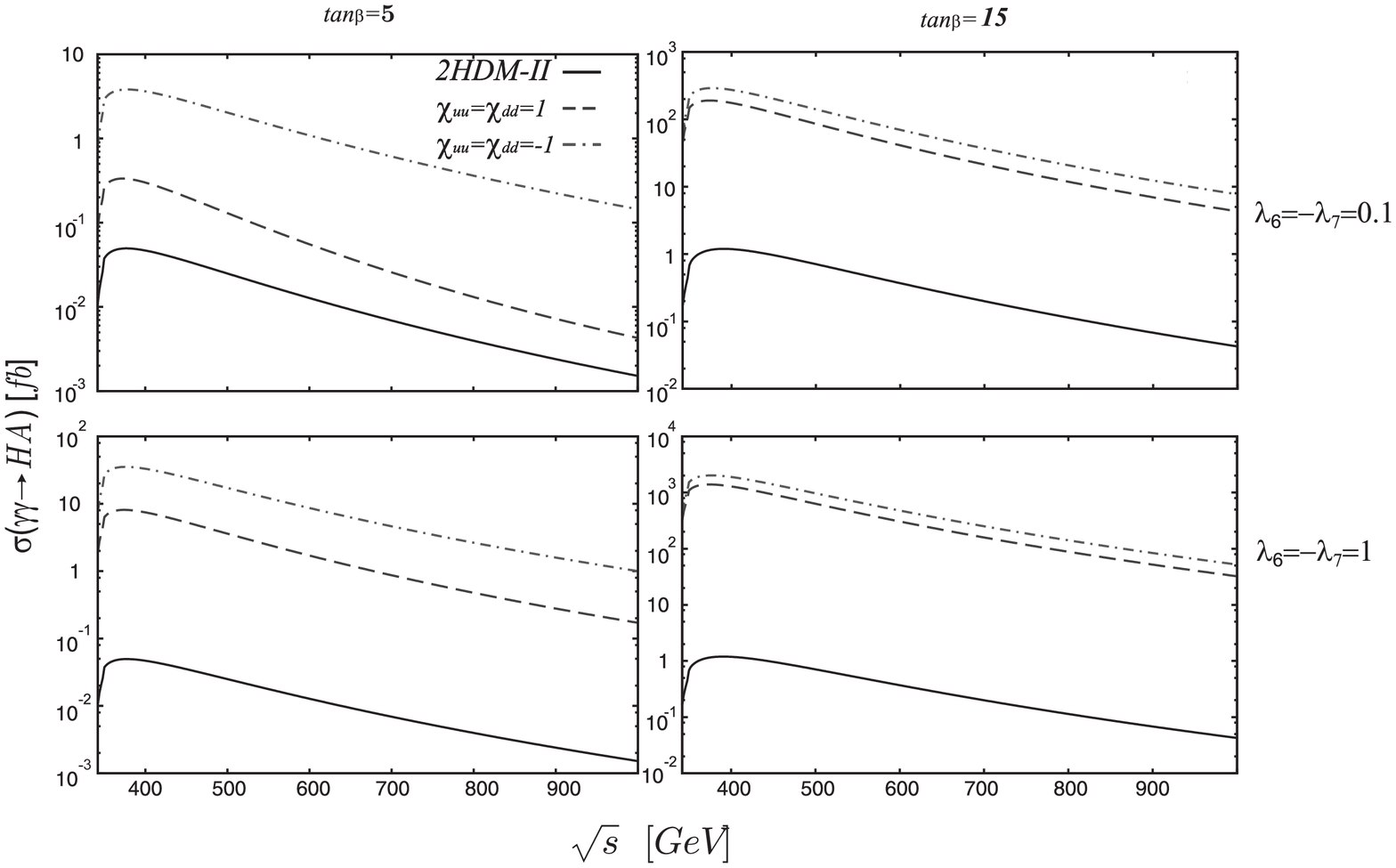}
\caption{Behavior of the cross section for the process
$\gamma\gamma\to HA$ as a function of  the center-of-mass
energy $\sqrt{s}$ in the Scenario II (SM-like). The description of the plots is the same as in Figure \ref{sml-2}.  }
\label{sml-6}
\end{figure}

\subsection{ Scenario III (a more general case of 2HDM-III)}
As already commented, this scenario is much more general than the
scenario II, because  arbitrary couplings of neutral Higgs bosons
to SM particles are assumed. Also, the contributions of the Higgs
potential $\lambda_6$ and $\lambda_7$ parameters, as well as the
contributions of the Yukawa texture in the couplings $\phi f
\bar{f}$, are included. As commented at the end of Sec. \ref{model}, the degenerate and the nondegenerate cases, as well as the case with a light CP-odd scalar will be considered.

\subsubsection{The nondegenerate case}

We first discuss the nondegenerate case, defined by the values $m_{H^\pm}=400$ GeV, $m_A=350$ GeV, $M_H=520$ GeV, $\mu_{12}=120$ GeV, and $m_h=120$ GeV. The set of values $\lambda_7= -\lambda_6=-1$ and $\lambda_7= -\lambda_6=-0.1$ are considered. In addition, it is assumed that $\alpha=\beta$ and $\alpha=\beta\pm \pi/2$.

\begin{figure}[htb]
\centering
\includegraphics[width=6 in]{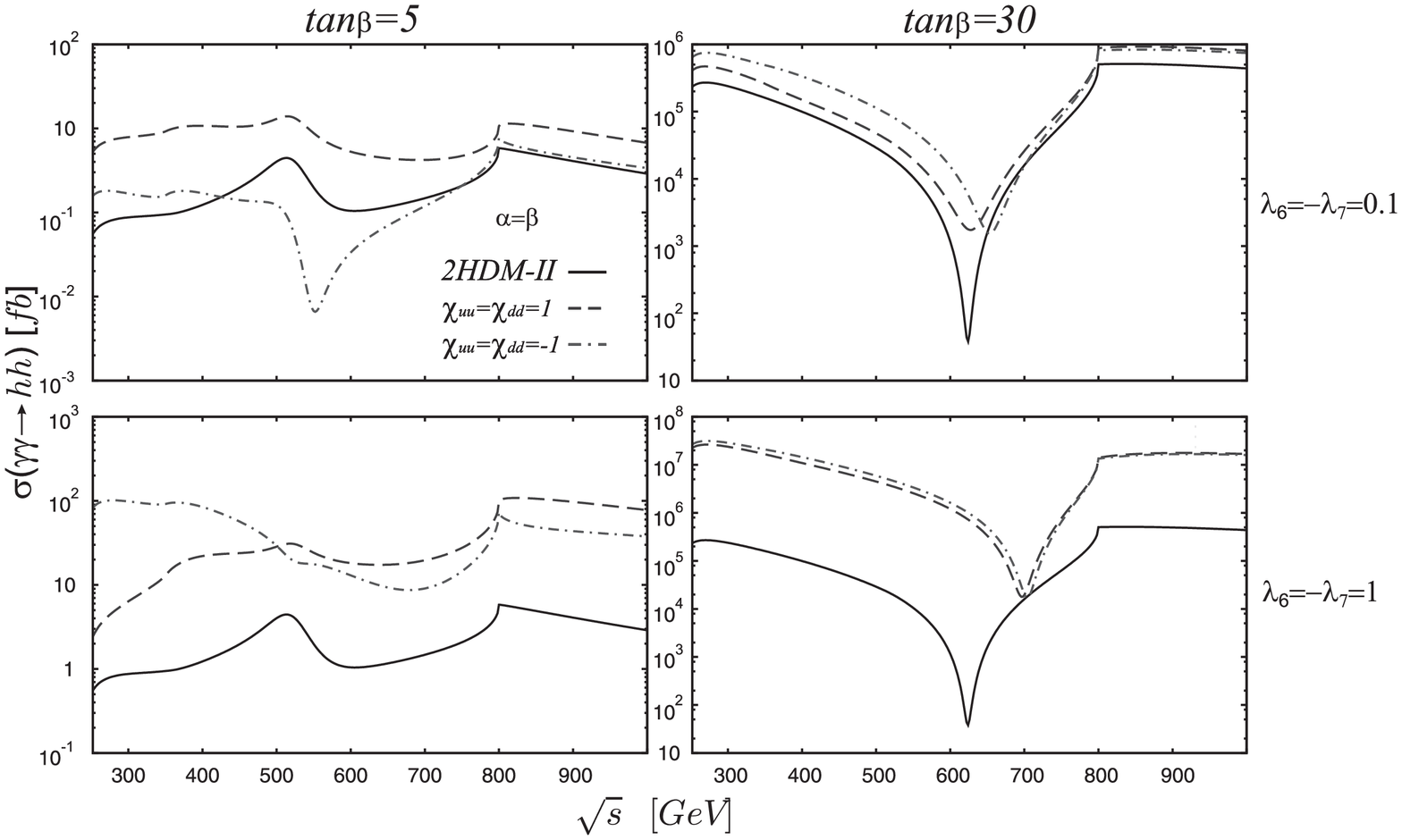}
\caption{Behavior of cross section for the process
$\gamma\gamma\to hh$ as a function of  the center-of-mass
energy $\sqrt{s}$ in the Scenario III (the nondegenerate case).  The lines correspond to
2HDM-II (solid line), 2HDM-III with $\chi_{uu,dd}=1$ (dashed line)  and 2HDM-III when $\chi_{uu,dd}=-1$ (dashed-dotted line), for  $\tan\beta =5$ (left), 30  (right) in the cases $\lambda_6=-\lambda_7=0.1$ (up panels) and $\lambda_6=-\lambda_7=1$ (down panels). We choice the  mixing angle  $\alpha = \beta$. }
\label{gen-1}
\end{figure}

In the down panels of Fig. \ref{gen-1}, we show the  cross section for the process
$\gamma\gamma\to hh$ as a function of  the center-of-mass energy
$\sqrt{s}$. An important difference can be appreciated between two
specific values of the Yukawa matrices with textures $\chi=1 $ and
$\chi =-1$. For $\tan \beta = 5$ and $\chi=-1$,  the cross section
could be up to one order of magnitude larger than for the case
$\chi =1$,  when $\sqrt{s} <350$ GeV. In this region, we can get $
\sigma(\gamma \gamma \to hh)\sim 1 \times 10^2$ fb for $\sqrt{s} $
around 370 GeV. The cross section predicted by the 2HDM-II is two
(one) order of magnitude lower than the 2HDM-III prediction with
$\chi =-1 (1)$. It can be seen that for $\sqrt{s} >500$ GeV and
$\chi=1 $ and $\chi =-1$, the cross section  is of the same order
of magnitude. However, in the 2HDM-III the cross section could be
larger than the result obtained in the context of the 2HDM-II, due
to the following choice  $\lambda_6=-\lambda_7=1$ of the
parameters. In the same figure, it can be appreciated a
spectacular enhancement of the cross section of $ \sigma(\gamma
\gamma \to hh)\sim 3 \times 10^7$ fb for $\tan \beta = 30$,
$\chi=-1$, and $\sqrt{s} =350$  GeV. The cross section predicted
by the 2HDM-II is one order of magnitude lower than the one
predicted by the 2HDM-III. On the other hand, it can be appreciated from the up panels of the same figure that in $\lambda_6=-\lambda_7=0.1$ case, the corresponding cross sections are suppressed by about one order of magnitude with respect to those obtained in the $\lambda_6=-\lambda_7=1$ case.

\begin{figure}[h]
\centering
\includegraphics[width=6 in]{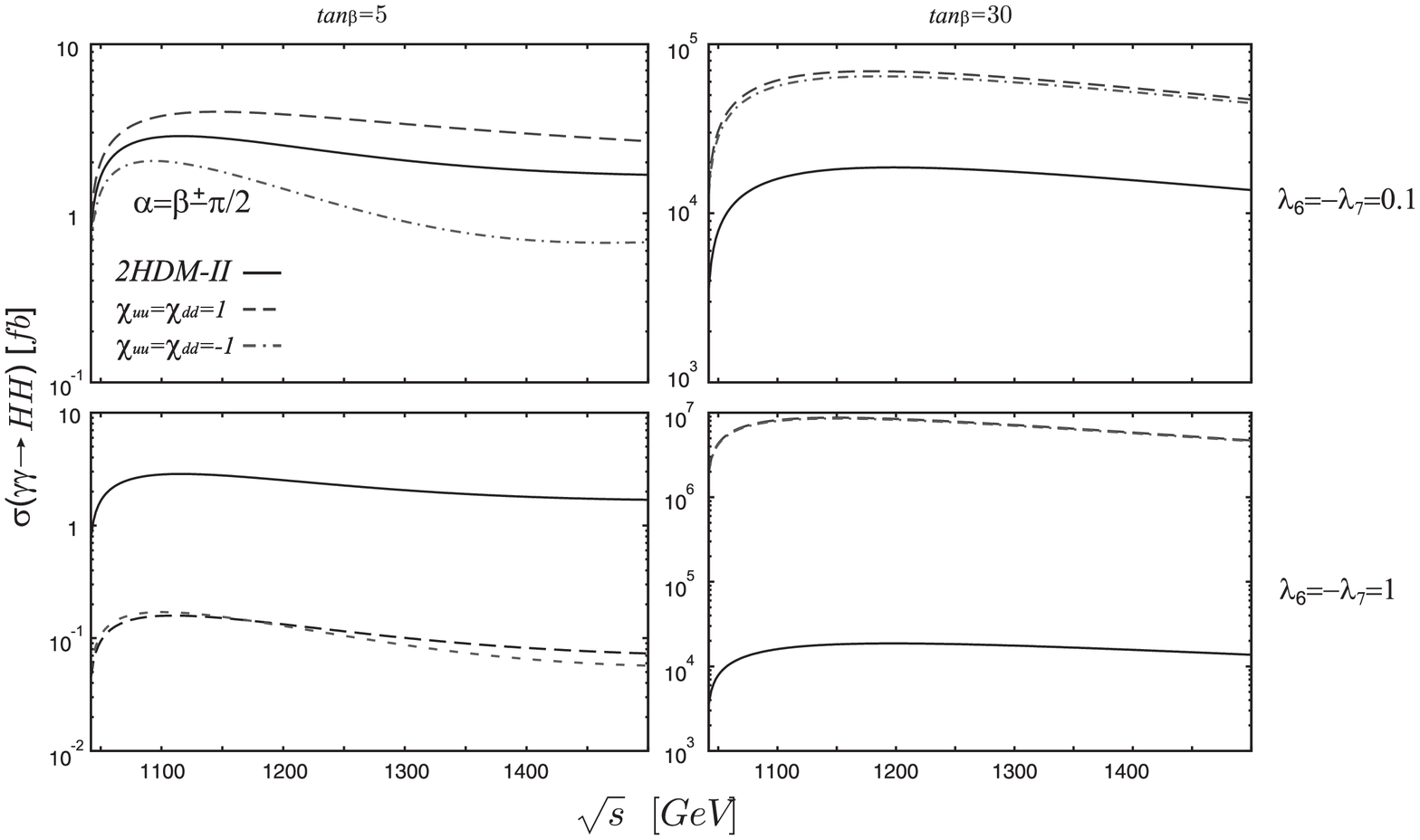}
\caption{Behavior of cross section for the process
$\gamma\gamma\to HH$ as a function of  the center-of-mass
energy $\sqrt{s}$ in the Scenario III, with a nondegenerate masses spectrum.  The lines correspond to
2HDM-II (solid line), 2HDM-III with $\chi_{uu,dd}=1$ (dashed line)  and 2HDM-III when $\chi_{uu,dd}=-1$ (dashed-dotted line), for  $\tan\beta =5$ (left), 30 (right) in the cases $\lambda_6=-\lambda_7=0.1$ (up panels) and $\lambda_6=-\lambda_7=1$ (down panels). We choice the  mixing angle  $\alpha = \beta \pm \pi/2$.}
\label{gen-2}
\end{figure}

On the other hand, the cross section for the $\gamma\gamma\to HH$ reaction is shown in Fig. \ref{gen-2} as a function of  the center-of-mass
energy $\sqrt{s}$. The signal for this process could be relevant in the TeVs region, $\sqrt{s}> 1$ TeV. Therefore, this mode could be far away of the reach of early linear colliders. It can be appreciated from this figure that the predictions of the THDM-III approximates to that of the THDM-II in the case $\lambda_6=-\lambda_7=0.1$.
\begin{figure}[h]
\centering
\includegraphics[width=6 in]{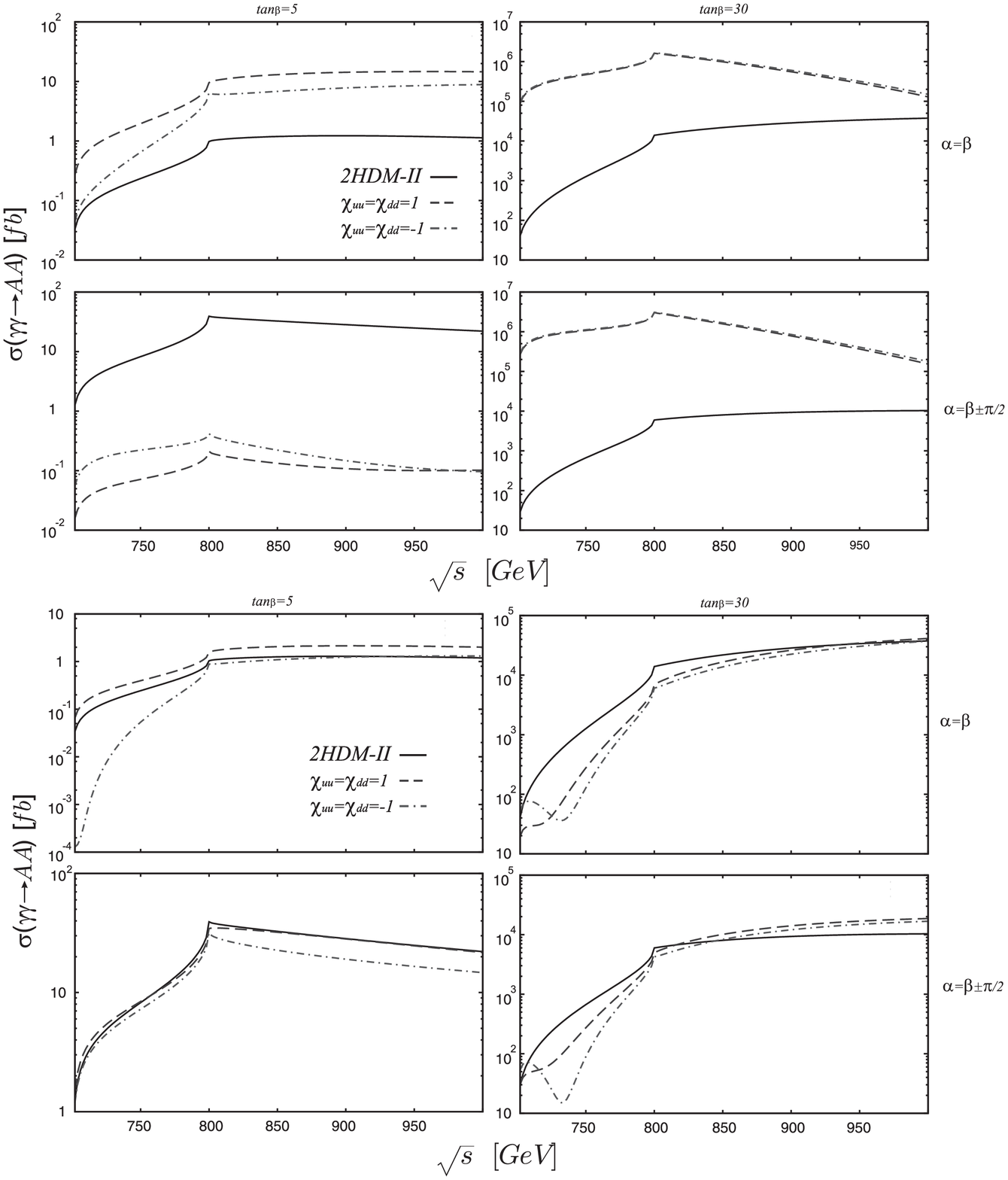}
\caption{Behavior of cross section for the process
$\gamma\gamma\to AA$ as a function of  the center-of-mass
energy $\sqrt{s}$ in the Scenario III, with a nondegenerate masses spectrum.  The lines correspond to
2HDM-II (solid line), 2HDM-III with $\chi_{uu,dd}=1$ (dashed line)  and 2HDM-III when $\chi_{uu,dd}=-1$ (dashed-dotted line), for  $\tan\beta =5$ (left), 30  (right), taking $\alpha = \beta$ (up panels) and $\alpha = \beta \pm \pi/2$ (down panels). The up(down) set of graphics corresponds to the case $\lambda_6=-\lambda_7=1$ ($\lambda_6=-\lambda_7=0.1$).}
\label{gen-3}
\end{figure}

As far as the $\gamma \gamma \to AA$ process is concerned, the
corresponding cross section as a function  of the center-of-mass
energy $\sqrt{s}$ is shown in Fig. \ref{gen-3}. This mode is
important in the 2HDM-III for center-of-mass energies above $700$
GeV and for large values of $\tan \beta$. From this figure, it can
be appreciated that in the case $\lambda_6=-\lambda_7=1$ (up set of graphics inf Fig. \ref{gen-3}), $ \sigma(\gamma \gamma \to AA)\sim 1 \times
10^6$ fb for $\sqrt{s} = 800$  GeV, $\tan \beta = 30 $, and
$\alpha = \beta $ or  $\alpha = \beta \pm \pi/2$. The 2HDM-III
predictions are two orders of magnitude larger than the ones of
2HDM-II. In the scenario with $\tan \beta =5 $, the cross section
predicted by 2HDM-III is of order of 10 fb for $\sqrt{s}=800$,
$\chi= 1$, and $\alpha = \beta $ , whereas the 2HDM-II prediction
is about one order of magnitude lower. However, the situation
changes drastically when $\alpha = \beta \pm \pi/2 $, as in this
case the 2HDM-II contribution dominates. On the other hand, it can be appreciated from these figures that in the case $\lambda_6=-\lambda_7=0.1$ (down set of graphics in Fig. \ref{gen-3}), the corresponding cross sections are of the same order of magnitude that those predicted by the THDM-II.

\begin{figure}[h]
\centering
\includegraphics[width=6 in]{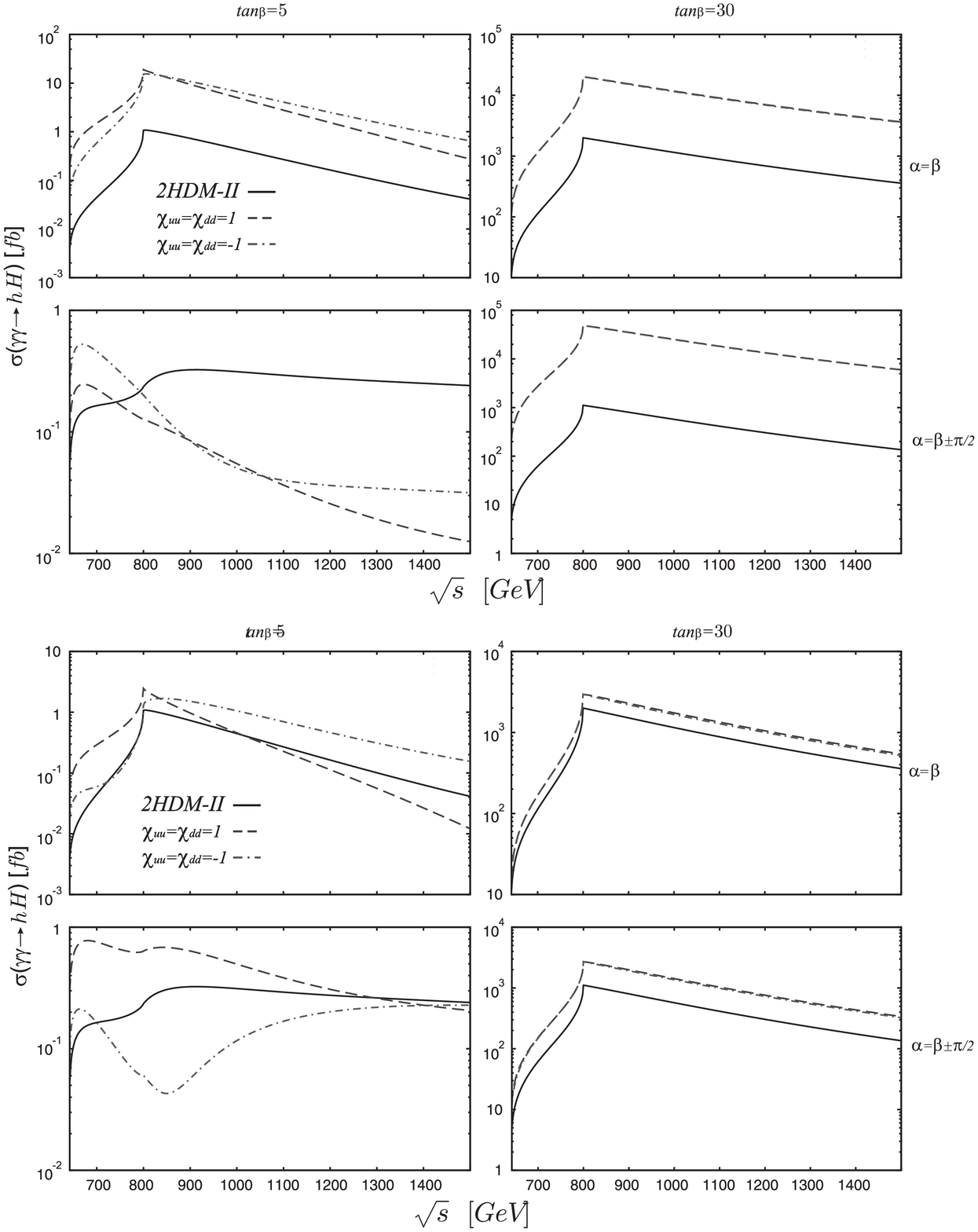}
\caption{The figure shows the behavior of cross section for the process
$\gamma\gamma\to hH$ as a function of  the center-of-mass
energy $\sqrt{s}$ in the Scenario III for the nondegenerate case.  The description of the plots is the same as in Figure \ref{gen-3}.}
\label{gen-4}
\end{figure}

We now turn to discuss the process $\gamma\gamma\to hH$. The corresponding cross section as a function  of the center-of-mass energy $\sqrt{s}$ is shown in Fig. \ref{gen-4}, in which the up set of graphics corresponds to the case $\lambda_6=-\lambda_7=1$, whereas the down set was obtained using the values  $\lambda_6=-\lambda_7=0.1$. It is found that this process is sensitive to $\tan \beta$ and the mixing angle $\alpha$. The cross section can reach a value of 20 fb for $\tan \beta =5$ and  $\alpha=\beta$. For large $\tan \beta$ values, the cross section is enhanced by several orders of magnitude. In fact, $ \sigma(\gamma \gamma \to hH)\sim 1 (5)\times 10^4$ fb for $\sqrt{s} = 800$  GeV, $\tan \beta = 30 $, $\alpha = \beta $ ($\alpha = \beta \pm \pi/2$), and  $\chi = \pm 1$. From these figures, it can be appreciated that the prediction of the THDM-III for the  $\lambda_6=-\lambda_7=0.1$ case is about one order of magnitude lower than that for the  $\lambda_6=-\lambda_7=1$ case, and clearly tends to the THDM-II prediction.

\begin{figure}[h]
\centering
\includegraphics[width=6 in]{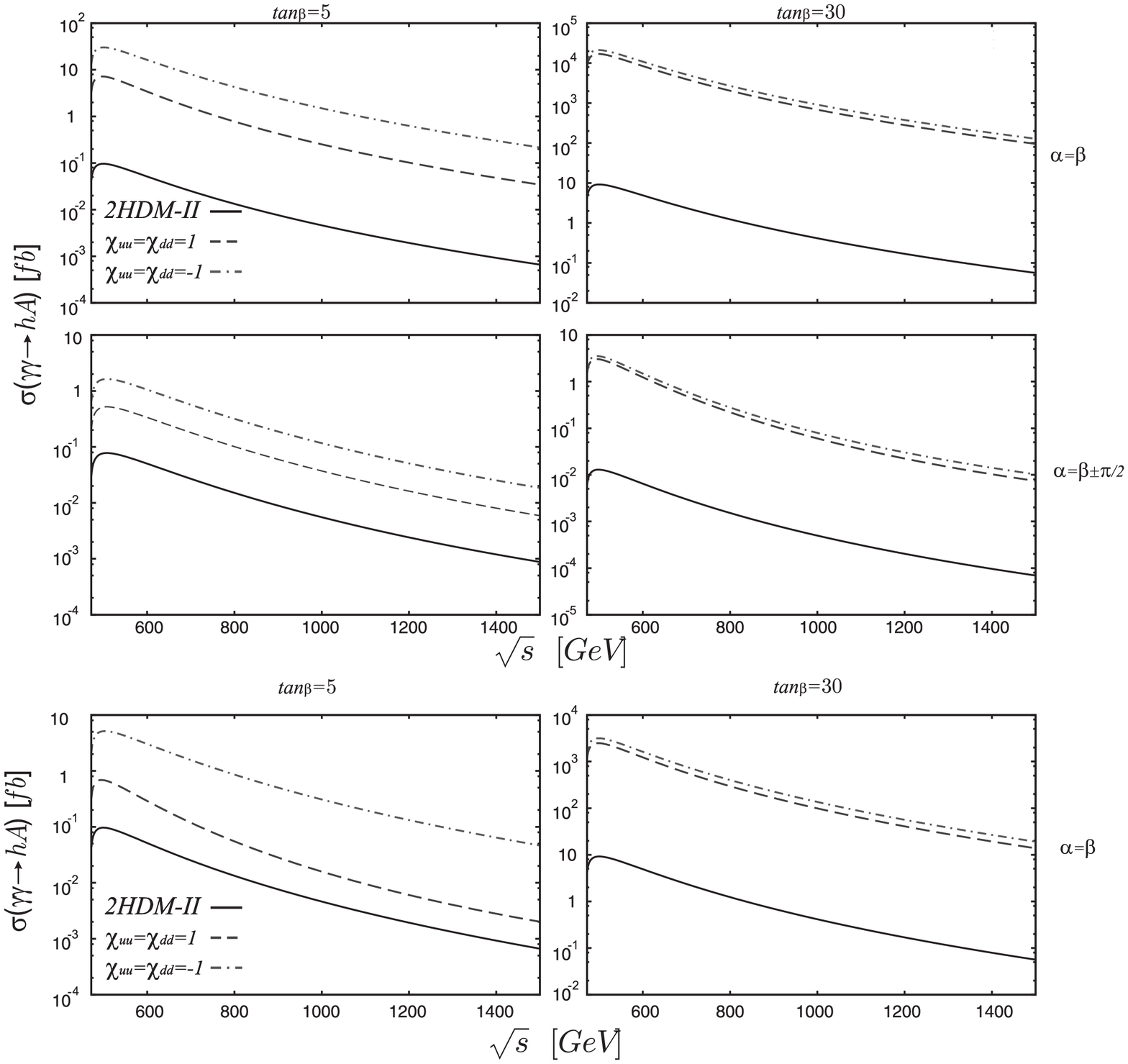}
\caption{The figure shows the behavior of cross section for the process
$\gamma\gamma\to hA$ as a function of  the center-of-mass
energy $\sqrt{s}$ in the Scenario III for the nondegenerate case. The description of the plots is the same as in Figure \ref{gen-3}. In the down set of diagrams, only the case $\alpha=\beta$ is considered.}
\label{gen-5}
\end{figure}

The cross section for the $\gamma \gamma \to hA$ as a function  of the center-of-mass energy $\sqrt{s}$ is shown in Fig. \ref{gen-5}, in which the up set of graphics corresponds to the $\lambda_6=-\lambda_7=1$ case, whereas the down set of figures arises from the $\lambda_6=-\lambda_7=0.1$ case. This cross section is quite sensitive to the mixing angle $\alpha$. A relevant value for the cross section arises when  $\tan \beta $ is large  and $\alpha = \beta$. In fact, $ \sigma(\gamma \gamma \to hA)\sim 1\times 10^4$ fb for $\sqrt{s} = 500$  GeV,  $\tan \beta = 30 $, and $\alpha = \beta $. It is important to notice that in the case  $\lambda_6=-\lambda_7=0.1$, only results for $\alpha=\beta$ are presented, as for $\alpha=\beta \pm \pi/2$ the cross sections are essentially independent on the $\lambda_6$ and $\lambda_7$ parameters, being therefore almost identical to those of the down panel of the up set of graphics. It can be appreciated from these figures that the cross sections for the  $\lambda_6=-\lambda_7=0.1$ case are about one order of magnitude lower than those for the  $\lambda_6=-\lambda_7=1$ case.

\begin{figure}[h]
\centering
\includegraphics[width=5 in]{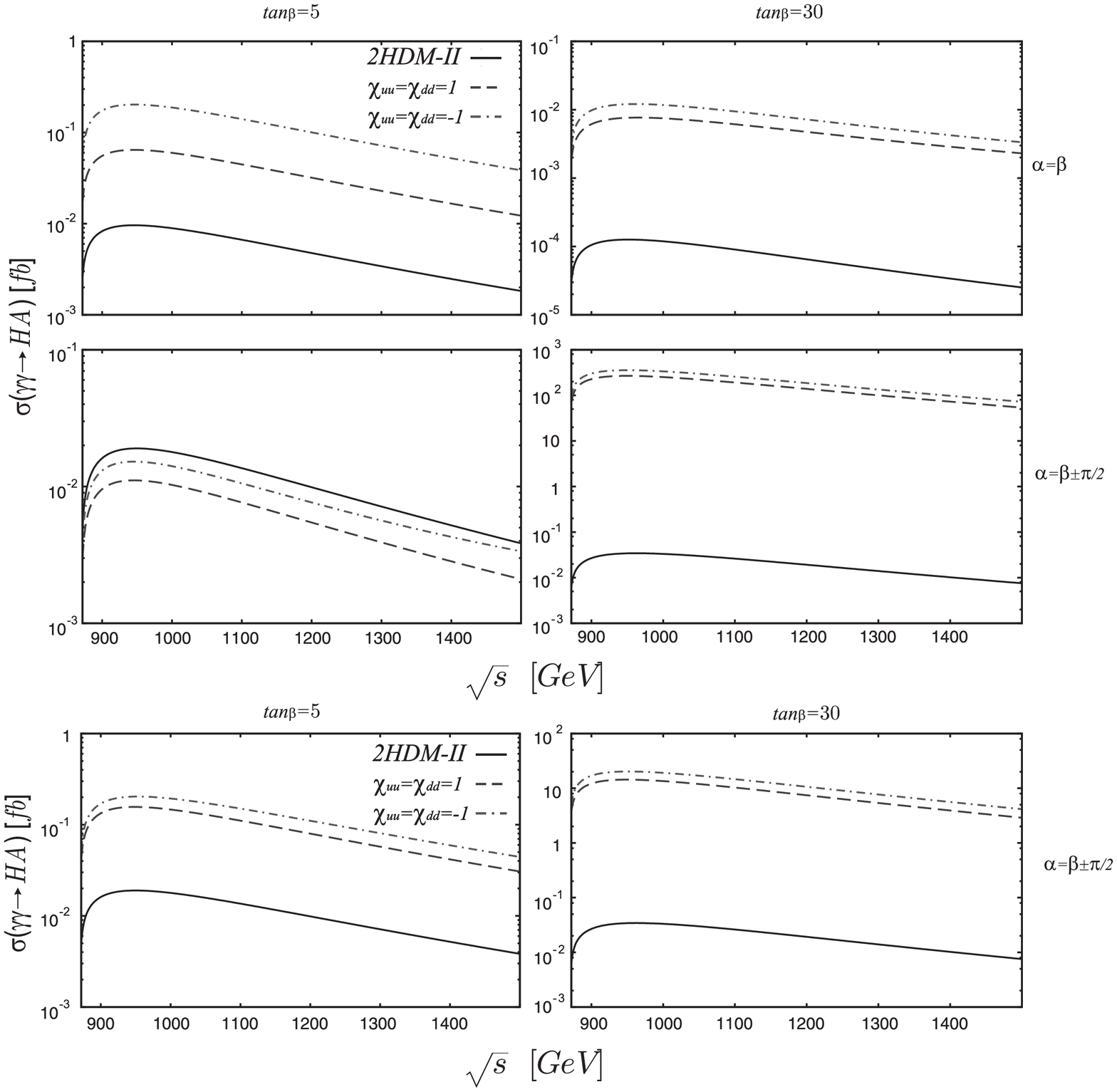}
\caption{Behavior of cross section for the process
$\gamma\gamma\to HA$ as a function of  the center-of-mass
energy $\sqrt{s}$ in the Scenario III for the nondegenerate case. The description of the plots is the same as in Figure \ref{gen-3}. In the down set of diagrams, only the case $\alpha=\beta \pm \pi/2 $ is considered.}
\label{gen-6}
\end{figure}

The cross section for the $\gamma\gamma\to HA$ process is shown in Fig. \ref{gen-6} as a function of the center--mass energy, in which the up set of graphics corresponds to the $\lambda_6=-\lambda_7=1$ case, whereas the down set of figures arises from assuming $\lambda_6=-\lambda_7=0.1$. Relevant cross sections are predicted by the 2HDM-III of order $10^2$ fb for $\tan \beta = 30$, $\chi =  \pm 1$, and energies around $900$ GeV. It is important to notice that in the case  $\lambda_6=-\lambda_7=0.1$, only results for $\alpha=\beta+\pi/2$ are presented, as for $\alpha=\beta$ the cross sections are essentially independent on the $\lambda_6$ and $\lambda_7$ parameters, being therefore almost identical to those of the down panel of the up set of graphics. It can be appreciated from this figure that for $\tan \beta=5$ the cross section in the $\lambda_6=-\lambda_7=0.1$ case is one order of magnitude larger than in the  $\lambda_6=-\lambda_7=0.1$ case, in contrast with the behavior observed in all the other processes.

\subsubsection{The degenerate case}
In this paragraph, we present results for the degenerate case, which is defined in Sec. II. Only the case $\lambda_6=-\lambda_7=1$ will be considered, as the $\lambda_6=-\lambda_7=0.1$ case leads to cross sections suppressed by about one order of magnitude with respect to the former one. Although in general terms the cross sections for $\tan \beta=30$ tend to be about two orders of magnitude larger than those obtained using $\tan \beta=5$, we have preferred to make predictions using only the latter value because in this case the predictions of the two versions of the model (THDM-III and THDM-II) can be clearly distinguished.

\begin{figure}[h]
\centering
\includegraphics[width=5 in]{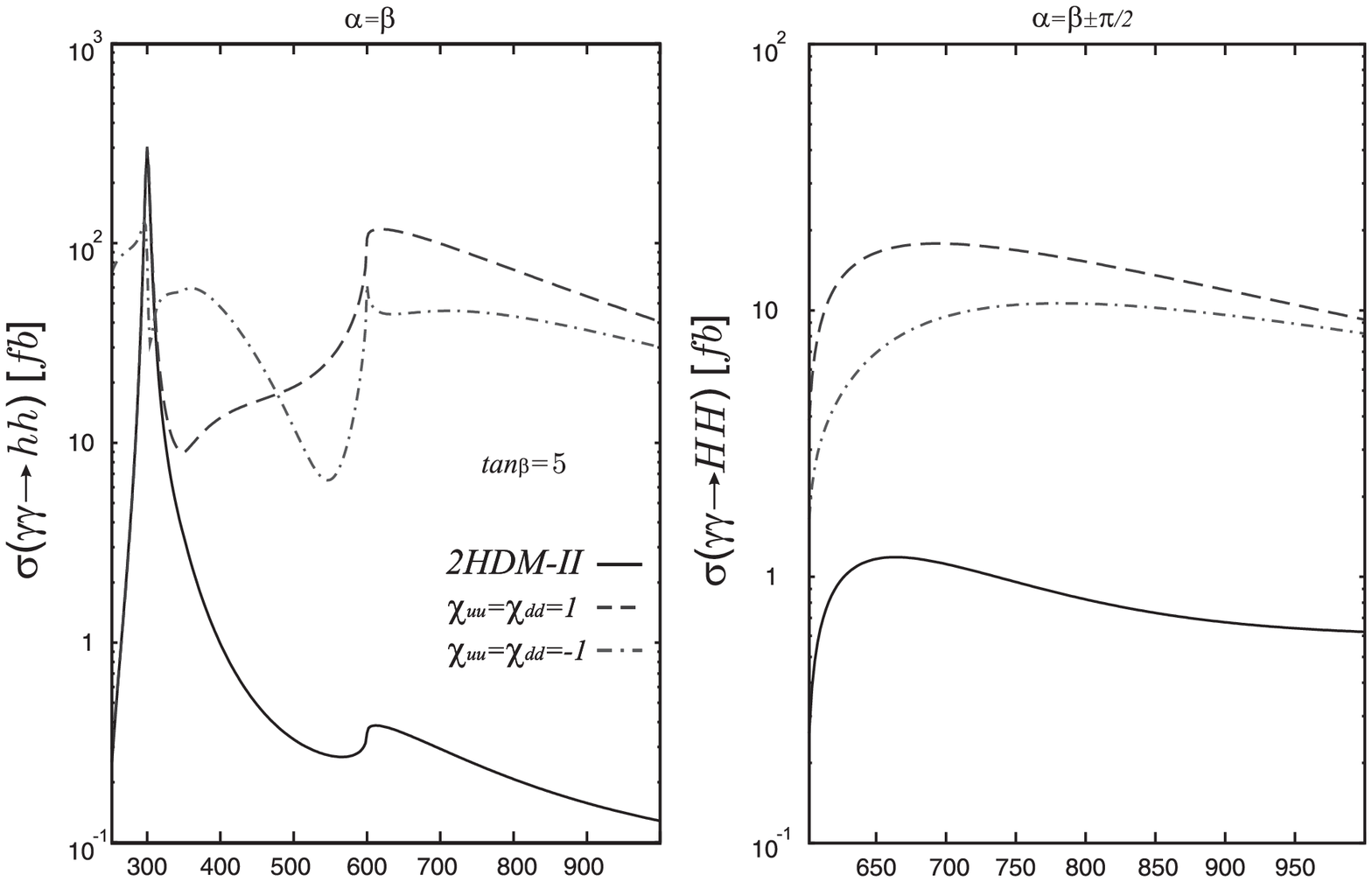}
\caption{Behavior of cross section for the processes
$\gamma\gamma\to hh$ (left) and $\gamma\gamma\to HH$ (right) as a function of  the center-of-mass
energy $\sqrt{s}$ in the Scenario III for the degenerate case.}
\label{deg-1}
\end{figure}

In Fig. \ref{deg-1}, the behavior of the cross sections for the processes $\gamma \gamma \to hh$ and $\gamma \gamma \to HH$ are shown as a function of the center--mass--energy, with $\alpha=\beta$ in the former process and $\alpha=\beta \pm \pi/2$ in the latter one. Besides to optimize the cross sections, these choice of values maximize the differences between both models. It can be appreciated from this figures that the THDM-III predicts cross sections as large as $10^2 \, fb$ and $10\, fb$ for the $hh$ and $HH$ channels, respectively, which are two and one orders of magnitude larger than those predicted by the THDM-II.

\begin{figure}[h]
\centering
\includegraphics[width=5 in]{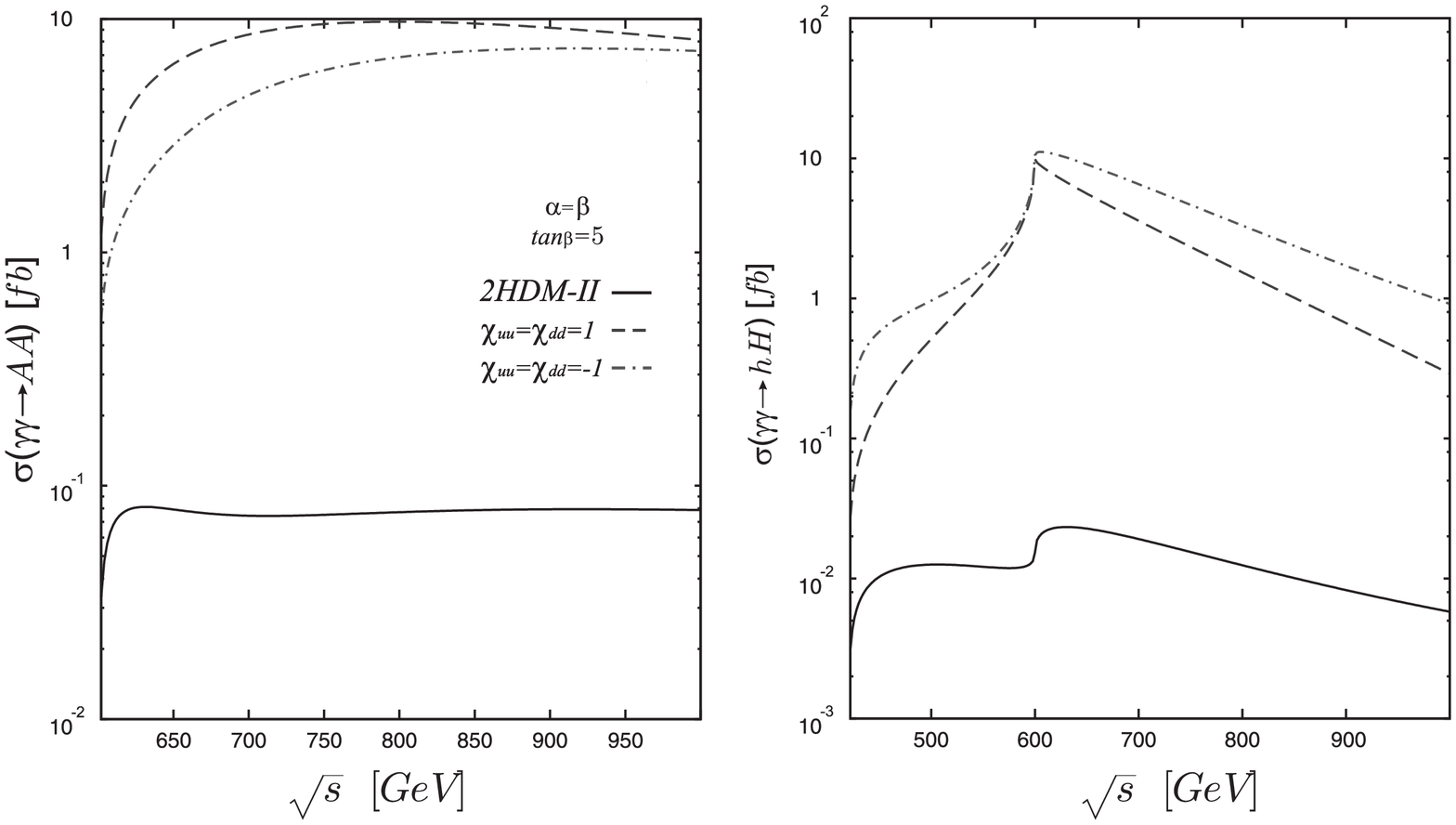}
\caption{Behavior of cross section for the processes
$\gamma\gamma\to AA$ (left) and $\gamma\gamma\to hH$ (right) as a function of  the center-of-mass
energy $\sqrt{s}$ in the Scenario III for the degenerate case.}
\label{deg-2}
\end{figure}

The cross sections for the processes $\gamma \gamma \to AA$ and $\gamma \gamma \to hH$ are shown in Fig. \ref{deg-2} as a functions of the center--mass--energy, for $\alpha=\beta$ in both cases. From this figure, it can be appreciated that the THDM-III prediction for the cross sections of both processes range from about $1\, fb$ to $10\, fb$ in the energies range shown. In contrast, the THDM-II predict cross sections quite suppressed ($10^{-1}\, fb$ for  $\gamma \gamma \to AA$ and $10^{-2}\, fb$ for  $\gamma \gamma \to hH$), which varies slightly in all the energies range considered.

\begin{figure}[h]
\centering
\includegraphics[width=5 in]{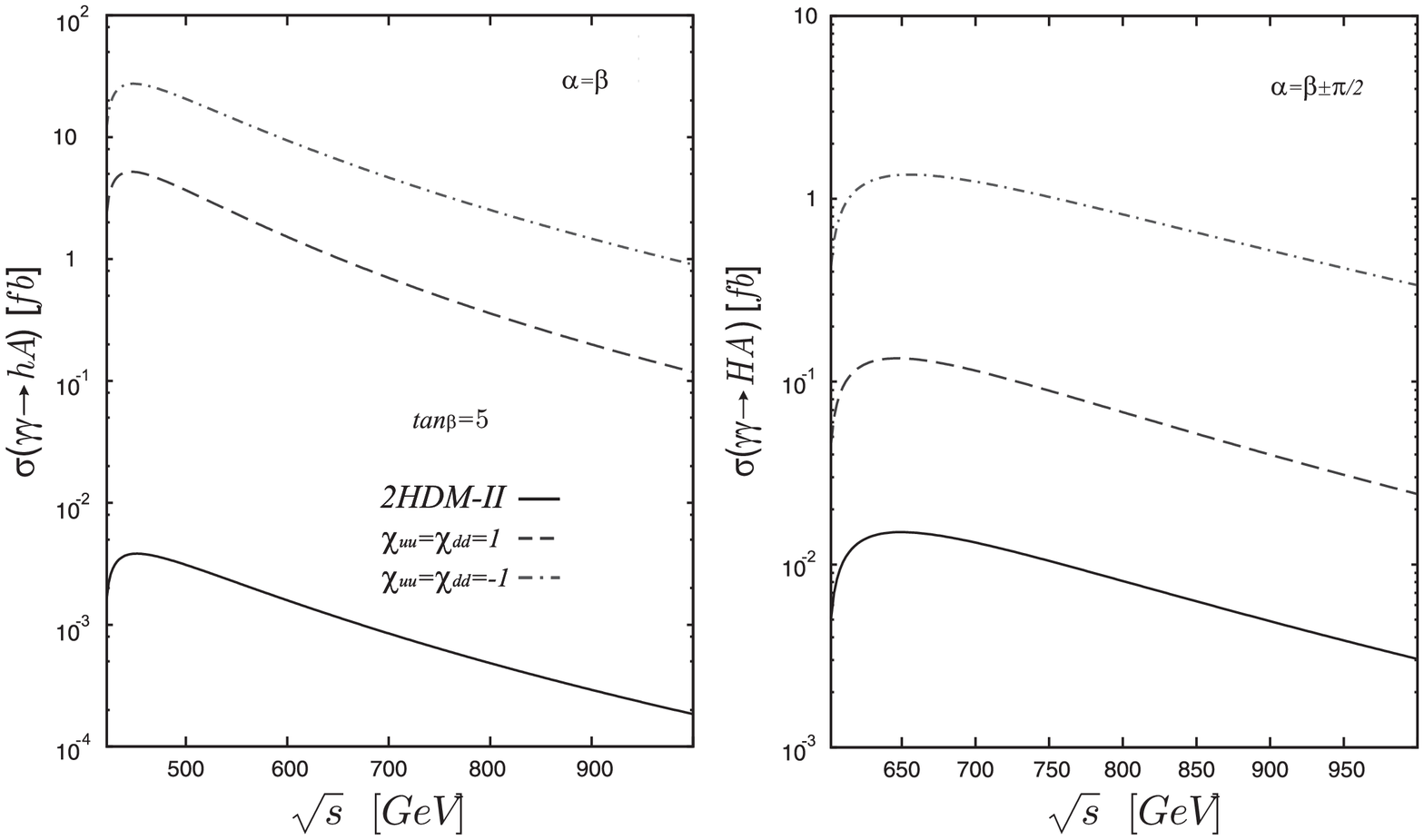}
\caption{Behavior of cross section for the processes
$\gamma\gamma\to hA$ (left) and $\gamma\gamma\to HA$ (right) as a function of  the center-of-mass
energy $\sqrt{s}$ in the Scenario III for the degenerate case.}
\label{deg-3}
\end{figure}

The cross sections for the processes $\gamma \gamma \to hA$ and $\gamma \gamma \to HA$ are shown in Fig. \ref{deg-3} as a function of the center--mass--energy. The relations used between the $\alpha$ and $\beta$ angles are shown in the figure. In this case, the predictions of the THDM-III ranges from $1\, fb$ to $10\, fb$ for the $\gamma \gamma \to hA$ process, whereas the prediction for the $\gamma \gamma \to HA$ reactions is one order of magnitude lower. The predictions of the THDM-II are quite suppressed, as both cross sections are of order of $10^{-2}\, fb$ or lower.

\subsubsection{A light CP--odd scalar}
\begin{figure}[h]
\centering
\includegraphics[width=5 in]{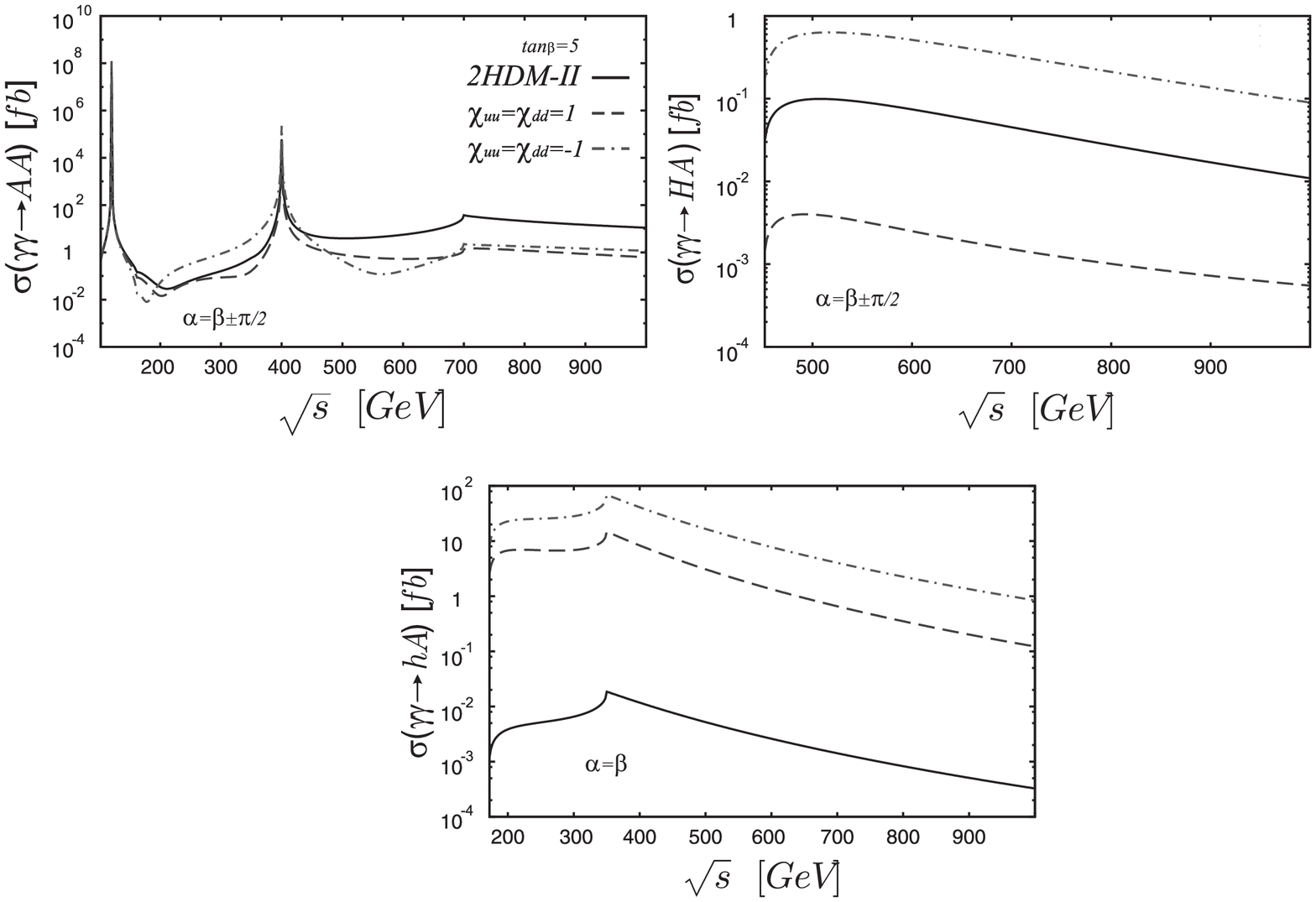}
\caption{Behavior of cross section for the processes
$\gamma\gamma\to AA$, $\gamma\gamma\to HA$, and $\gamma\gamma\to HA$ as a function of  the center-of-mass
energy $\sqrt{s}$ in the scenario of a light CP--odd scalar.}
\label{light-A}
\end{figure}

In this paragraph, we discuss the very interesting case of a light CP--odd scalar $A$, which is allowed by the current constraints on the model. We consider the three possible processes, namely, $\gamma \gamma \to AA$, $\gamma \gamma \to HA$, and $\gamma \gamma \to hA$. In Fig. \ref{light-A}, the behavior of the cross sections for these processes as a functions of the center--mass--energy is shown, in a scenario with $m_A=50$ GeV, $m_h=120$ GeV, $m_{H^\pm}=350$ GeV, $m_H=400$ GeV, $\mu_{12}=70$ GeV, and $\tan \beta=5$. The value $\alpha=\beta \pm \pi/2$ for the $\gamma \gamma \to AA$ and $\gamma \gamma \to HA$ processes is assumed, whereas in the case of the  $\gamma \gamma \to hA$ reaction it is assumed that $\alpha=\beta$. In all these processes, it is assumed that $\lambda_6=-\lambda_7=1$ and that $\chi_{uu}=\chi_{dd}=\{1,-1\}$. Our notation and conventions are shown in the first graph of Fig. \ref{light-A}. From the first graph of this figure, it can be appreciated three resonant effects for the $\gamma \gamma \to AA$ process, centered at energies $\sqrt{s}=120 \, GeV =m_h$,  $\sqrt{s}=400 \, GeV =m_H$, and  $\sqrt{s}=350 \, GeV =2m_t$. The resonant effects due to $m_h$ and $m_H$ are spectacular, as the cross section can reach values of up to $10^8 \, fb$ and $10^{6}\, fb$, respectively. The resonant effect at $2m_t$ is less significative, as it occurs through a 1--loop fluctuation. Apart from these resonant effects, the values of the cross section are within the range of variation encountered in other scenarios analyzed previously, as the cross section predicted by the THDM-III ranges approximately from $10^{-1}\, fb$ to $1\, fb$ in both the $\chi_{uu}=\chi_{dd}=-1$ and $\chi_{uu}=\chi_{dd}=1$ scenarios, whereas in the THDM-II the corresponding cross section is about one order of magnitude larger. As far as the $\gamma \gamma \to HA$ process is concerned, it can be seen from this figure that the cross section predicted by the THDM-III ranges from $10^{-1}\, fb$ to $1\, fb$ for the scenario with $\chi_{uu}=\chi_{dd}=-1$, whereas for $\chi_{uu}=\chi_{dd}=1$ the cross section ranges from $10^{-3}\, fb$ to $10^{-2}\, fb$. The corresponding cross section predicted by the THDM-II ranges from $10^{-2} \, fb$ to $10^{-1} \, fb$ in the same range of variation of the center--mass--energy. Finally, it can be appreciated from the third graph of Fig. \ref{light-A} that the cross section for the $\gamma \gamma \to hA$ reaction ranges from $10\, fb$ to $10^{2} \, fb$ in the THDM-III in the scenario $\chi_{uu}=\chi_{dd}=-1$, and from $ 1\, fb$ to $10\, fb$ in the scenario $\chi_{uu}=\chi_{dd}=-1$. The corresponding cross section in the THDM-II is quite suppressed, as it ranges from $10^{-4}\, fb$ to $10^{-2}\, fb$ in the same domain of energies. The cross sections for all these processes can be enhanced by at least in one order of magnitude if $\tan \beta=30$ is used instead of $\tan \beta=5$.

\section{Summary}
\label{conclusions} In this paper, a comprehensive study of the
one--loop $\gamma \gamma \to \phi_i \phi_j$ ($\phi_i=h, H, A$)
processes in the context of a general version of the two Higgs
doublet model (2HDM-III) was presented. A nonlinear
$R_\xi$--gauge, which allows us to define the $W$ gauge boson
propagator in a covariant way under the electromagnetic gauge
group, was used. This gauge, which reduces significantly the
number of Feynman diagrams to be considered in comparison with
those that must be calculated in conventional linear gauges, makes
the issues of electromagnetic gauge invariance and cancelation of
divergencies simpler. Explicit analytical expressions for all the
possible modes, namely $hh$, $HH$, $AA$, $hH$, $hA$, and $HA$,
were presented. The corresponding amplitudes are completely
general in the sense that they can be used for any version of the
2HDM. The version of the 2HDM that is considered in this work,
which we called simply 2HDM-III, comprise the implementation of a
flavor symmetry in the Yukawa sector, namely a four-zero Yukawa
Texture, which allows us to suppress FCNC effects without
necessity of introducing a discrete symmetry. Due to this, a Higgs
potential, more general than the one considered in the 2HDM-II
version, can be introduced. This Higgs potential includes two
dimensionless parameters, $\lambda_6$ and $\lambda_7$, to which
the cross sections for the $\gamma \gamma \to \phi_i \phi_j$
processes are quite sensitive. The $\gamma \gamma \to \phi_i
\phi_j$ mechanisms for Higgs pair production were analyzed in
three scenarios of the 2HDM-III. In the scenario I (the decoupling
case), the $\gamma \gamma \to hh$ reaction was studied in the
two possible cases in which the decoupling operates, namely, when
$\mu^2_{12}\gg v^2$ or $\mu^2_{12}\sim v^2$ but assuming that
$\tan \beta$ or $\cot \beta$ are large, depending on the
configuration chosen for the $\lambda_6$ and $\lambda_7$
parameters. In both cases, it is found that, in the heavy mass
limit of the charged Higgs, the cross section for this reaction
approach to the well known SM result. In the scenario II
(SM--like), it is assumed that the $hVV$ ($V=W,Z$), $hhh$, and
$hhhh$ couplings are nearly indistinguishable from the
corresponding ones of the SM, but the $hf\bar{f}$ couplings can
deviate significantly from their SM counterparts $h_{SM} f
\bar{f}$. It was found that some combinations of Yukawa textures
with the $\lambda_6$ and $\lambda_7$ parameters, together with
large $\tan \beta$ lead to experimentally interesting cross
sections. In the scenario III (a more general case of 2HDM-III),
besides considering the contributions of the Higgs potential
through the $\lambda_6$ and $\lambda_7$ parameters, as well as the
contributions of the Yukawa texture in the couplings $\phi f
\bar{f}$, arbitrary couplings of neutral Higgs bosons to SM
particles were assumed. In this scenario, the implications of a light CP--odd scalar was studied.

In general terms, we can conclude that the parameters of the Higgs
potential $\lambda_6$ and $\lambda_7$ considerably enhance the
cross sections for the $\gamma \gamma \to \phi_i \phi_j$
processes. In almost all cases, the results of 2HDM-III are two
orders of magnitude larger than those obtained from the 2HDM-II. A
considerable enhancement for the cross sections of the processes
$\gamma \gamma \to \phi_i \phi_j$ was observed in the regime of
large $\tan \beta$.

\appendix
\section{Feynman rules in the nonlinear gauge}
\label{a1}
We present the couplings for the most polular types of 2HDM (I,
II, y III). The vertex can be written as
\begin{equation}
g_{\phi_i\bar f f}=-\frac{igm_f}{2m_W}\left\{
\begin{array}{cc}
{\cal G}_{\phi_a\bar f f},&{\rm for}\ h, H\\
i\gamma^5{\cal G}_{A\bar f f},& {\rm for}\ A\\
\end{array}
\right.
\end{equation}
where ${\cal G}_{\phi_i \bar f f}$ are dimensionless functions
expressed in the table (\ref{tablaYukawa}).
\begin{table}[h]
\begin{center}
\begin{tabular}{|l|l|l|l|}
\hline
${\cal G}_{\phi_i \bar f f}$ &Type-I&Type-II&Type-III  \\
\hline
 ${\cal G}_{hll}$, ${\cal G}_{hdd}$&$\frac{c_\alpha}{s_\beta}$
 &$\frac{-s_\alpha}{c_\beta}$&$\frac{-s_\alpha}{c_\beta}+\frac{c_{\alpha-\beta}\chi_{dd}}{\sqrt{2}c_\beta}$\\
 \hline
 ${\cal G}_{huu}$&$\frac{c_\alpha}{s_\beta}$&$\frac{c_\alpha}{s_\beta}$
 &$\frac{c_\alpha}{s_\beta}-\frac{c_{\alpha-\beta}\chi_{uu}}{\sqrt{2}s_\beta}$\\
 \hline
 ${\cal G}_{Hll}$,${\cal G}_{Hdd}$&$\frac{s_\alpha}{s_\beta}$&$\frac{c_\alpha}{c_\beta}$
 &$\frac{c_\alpha}{c_\beta}+\frac{s_{\alpha-\beta}\chi_{dd}}{\sqrt{2}c_\beta}$\\
 \hline
 ${\cal G}_{Huu}$&$\frac{s_\alpha}{s_\beta}$&$\frac{s_\alpha}{s_\beta}$
 &$\frac{s_\alpha}{s_\beta}-\frac{s_{\alpha-\beta}\chi_{uu}}{\sqrt{2}s_\beta}$\\
 \hline
 ${\cal G}_{All}$,${\cal G}_{Add}$&$t_\beta^{-1}$&$-t_\beta$
 &$-t_\beta+\frac{\chi_{dd}}{\sqrt{2}c_\beta}$\\
 \hline
 ${\cal G}_{Auu}$&$-t_\beta^{-1}$&$-t_\beta^{-1}$&$-t_\beta^{-1}+\frac{\chi_{uu}}{\sqrt{2}s_\beta}$\\
 \hline
\end{tabular}
\end{center}
\caption{\label{tablaYukawa} Dimensionless function that define
the Yukawa couplings  \cite{HHG, ourthdm3a,DiazCruz:2009ek}.}
\end{table}
\subsection{Yang-Mills Couplings}
This sector is strongly affected by the gauge-fixing procedure
that we used. This couplings can be found in \cite{NLGTHDM}. Here,
we only present the Feynman rules associated with the vertices
used in this work, namely,
$A_{\eta}(k_1)W^+_\lambda(k_2)W^-_\rho(k_3)$ and $A_\alpha A_\beta
W^+_\lambda W^-_\rho$. The corresponding vertex functions are
given by
\begin{eqnarray}
-ie\Gamma_{\lambda \rho \eta}(k_1,k_2,k_3)&=&-ie\left((k_3-k_2)_\eta
g_{\lambda \rho}+(k_1-k_3-\frac{1}{\xi}k_2)_\lambda g_{\rho
\eta}+(k_2-k_1+\frac{1}{\xi}k_3)_\rho g_{\lambda \eta}\right), \\
-ie^2\Gamma_{\alpha \beta \lambda \rho}&=&-ie^2\left(2g_{\alpha
\beta}g_{\lambda \rho}-(1-\frac{1}{\xi})(g_{\alpha \lambda}g_{\beta
\rho}+g_{\alpha \rho}g_{\beta \lambda})\right).
\end{eqnarray}
\subsection{Scalar and Kinetic sector}
Now we present the couplings that contain scalar particles, in
particular the vertex functions  necessary   for our calculation.
The vertex functions between  scalar Higgs can be written as
following:
\begin{eqnarray}
g_{\phi_a\phi_b\phi_c}&=&\frac{-igm_W}{4}{\cal
G}_{\phi_a\phi_b\phi_c},\\
g_{\phi_i H^\pm H^\mp }&=&\frac{-igm_W}{4}{\cal G}_{\phi_iH^\pm
H^\mp},\\
g_{H^\pm H^\mp \phi_i\phi_j}&=&\frac{-ig^2}{16}{\cal G}_{H^\pm
H^\mp \phi_i\phi_j}\\
g_{H^\pm G_{W}^\mp A}&=&\pm \frac{g(m_{H^\pm}^2-m_A^2)}{2m_W}\\
g_{\phi_a AA}&=&\frac{-igm_W}{4}{\cal G}_{\phi_a AA}
\end{eqnarray}
The explicit form of this vertex function can be found in
\cite{Gunion}.

The couplings that involve a pair of of $W$-Goldstone boson was
modified by the gauge-fixing procedure. This couplings can be
expressed as:
\begin{eqnarray}
g_{G_W^\pm G_W^\mp \phi_i}&=&\frac{-ig(m_i^2+2\xi
m_W^2)}{2m_W}{\cal G}_{\phi_i WW},\\
g_{G_W^\pm G_W^\mp \phi_i \phi_j}&=&\frac{-ig^2}{16}{\cal
G}_{G_W^\pm G_W^\mp \phi_i \phi_j}.
\end{eqnarray}
The explicit form of this couplings have to be take of the
reference  \cite{NLGTHDM}.

On the other hand, the  couplings involving gauge and scalar
fields are present in the table \ref{TABLE1}. Here, the
dimensionless  functions ${\cal G}$ can be written as:
\begin{eqnarray}
{\cal G}_{h W W}=&-{\cal G}_{W^\pm H^\mp
H}=-{\cal G}_{ZAH}&=s_{(\beta-\alpha)},\\
{\cal G}_{H W W}=&{\cal G}_{W^\pm H^\mp h}={\cal
G}_{ZAh}&=c_{(\beta-\alpha)}.
\end{eqnarray}

\begin{table}
\caption{\label{TABLE1} Structure of the couplings involving gauge
and scalar fields.}
\begin{ruledtabular}
\begin{tabular}{|l|l|l|l|}
\hline
Coupling & Vertex Function & Coupling & Vertex Function \\
\hline
 $g_{A_\mu H^-(p_-)H^+(p_+)}$ & $ie(p_--p_+)_\mu$ & $g_{A_\mu A_\nu
H^-H^+}$ & $2ie^2g_{\mu \nu}$ \\
\hline $g_{A_\mu G^-_W(p_-)G^+_W(p_+)}$ & $ie(p_--p_+)_\mu$ &
$g_{A_\mu
A_\nu G^-_WG^+_W}$ & $2ie^2g_{\mu \nu}$ \\
\hline $g_{\phi_aW^-_\mu W^+_\nu}$ & $igm_W{\cal
G}_{\phi_aWW}g_{\mu \nu}$ & $g_{(hh,HH,AA)W^-_\mu
W^+_\nu}$ & $\frac{ig^2}{2}g_{\mu \nu}$\\
\hline $g_{\phi_a(p_a)W^\pm_\mu H^\mp(p_{H^\pm})}$ & $\pm
\frac{ig}{2}{\cal G}_{W^\pm H^\mp \phi_a}(p_{H^\pm}-p_a)_\mu$ &
$g_{A_\mu W^\pm_\nu H^\mp \phi_a}$ & $\frac{ieg}{2}{\cal G}_{W^\pm H^\mp \phi_a}g_{\mu \nu}$\\
\hline $g_{A(p_A)W^\pm_\mu H^\mp(p_{H^\pm})}$ & $-
\frac{g}{2}(p_{H^\pm}-p_A)_\mu$ &
$g_{A_\mu W^\pm_\nu H^\mp A}$ & $\mp \frac{eg}{2}g_{\mu \nu}$\\
\hline $g_{\phi_a(p)W_\mu^{\pm}G_W^{\mp}}$&$\mp i g
p_{\mu}{\cal G}_{\phi_aWW}$&$g_{ZA(p_A)\phi_a(p_a)}$&$\frac{g}{2c_W}(p_{a}-p_A)_\mu{\cal G}_{ZA\phi_a}$\\

\end{tabular}
\end{ruledtabular}
\end{table}
\section{The scalar functions}
\label{a2} In this appendix we present the $C_0(a)$ and $D_0(a)$
scalar functions appearing in the amplitudes $A_{15}$ and $A_{25}$
associated with the processes $\gamma \gamma \to AA$ and $\gamma
\gamma \to \phi_a \phi_b$.

\begin{eqnarray}
&&C_0(1)=C_0(s,m^2_i,m^2_j,m^2_W,m^2_W,m^2_{H^\pm}), \\
&&C_0(2)=C_0(s,m^2_i,m^2_j,m^2_{H^\pm},m^2_{H^\pm},m^2_W), \\
&&C_0(3)=C_0(0,0,s,m^2_{H^\pm},m^2_{H^\pm},m^2_{H^\pm}), \\
&&C_0(4)=C_0(0,0,s,m^2_{W},m^2_W,m^2_W), \\
&&C_0(5)=C_0(0,m^2_i,u,m^2_{H^\pm},m^2_{H^\pm},m^2_{W}), \\
&&C_0(6)=C_0(0,m^2_i,u,m^2_{W},m^2_{W},m^2_{H^\pm}), \\
&&C_0(7)=C_0(0,m^2_i,t,m^2_{H^\pm},m^2_{H^\pm},m^2_{W}), \\
&&C_0(8)=C_0(0,m^2_i,t,m^2_{W},m^2_{W},m^2_{H^\pm}),\\
&&C_0(9)=C_0(0,t,m^2_j,m^2_{H^\pm},m^2_{H^\pm},m^2_W), \\
&&C_0(10)=C_0(0,t,m^2_j,m^2_{W},m^2_{W},m^2_{H^\pm}), \\
&&C_0(11)=C_0(0,u,m^2_j,m^2_{H^\pm},m^2_{H^\pm},m^2_W), \\
&&C_0(12)=C_0(0,u,m^2_j,m^2_{W},m^2_{W},m^2_{H^\pm}),
\end{eqnarray}
\begin{eqnarray}
&&D_0(1)=D_0(0,0,m^2_i,m^2_j,s,u,m^2_{H^\pm},m^2_{H^\pm},m^2_{H^\pm},m^2_W),
\\
&&D_0(2)=D_0(0,0,m^2_i,m^2_j,s,u,m^2_{W},m^2_{W},m^2_{W},m^2_{H^\pm}),
\\
&&D_0(3)=D_0(0,0,m^2_i,m^2_j,s,t,m^2_{H^\pm},m^2_{H^\pm},m^2_{H^\pm},m^2_W),
\\
&&D_0(4)=D_0(0,0,m^2_i,m^2_j,s,t,m^2_{W},m^2_{W},m^2_{W},m^2_{H^\pm}),
\\
&&D_0(5)=D_0(0,m^2_i,0,m^2_j,u,t,m^2_{H^\pm},m^2_{H^\pm},m^2_{W},m^2_{W}),
\\
&&D_0(6)=D_0(0,m^2_i,0,m^2_j,u,t,m^2_{W},m^2_{W},m^2_{H^\pm},m^2_{H^\pm}),
\end{eqnarray}
where $m_i=m_j=m_A$  for $\gamma\gamma\to AA$, and $m_i=m_a$ and
$m_j=m_b$  for $\gamma\gamma\to \phi_a\phi_b$.

\section*{Acknowledgments}

We acknowledge support by Conacyt, SNI, and Red-FAE (Mexico). JH-S and CGH acknowledge  R. Noriega-Papaqui for useful discussions.


\end{document}